\documentclass[12pt]{article}
\usepackage[left=2.3cm,right=2.3cm,top=2cm,bottom=2.5cm]{geometry}
\usepackage{amsfonts}
\usepackage{amsmath}
\usepackage{amsthm}
\usepackage{hyperref}
\usepackage{breqn}
\usepackage[usenames, dvipsnames]{color}
\allowdisplaybreaks 
\usepackage{graphicx}
\usepackage{caption}
\usepackage{subcaption}
\usepackage{float}
\usepackage{mathtools}
\usepackage{accents}
\usepackage{amssymb}

\providecommand{\keywords}[1]
{
	\small	
	\textbf{Keywords:} #1
}

\newtheorem{theorem}{Theorem}[section]
\newtheorem*{theorem*}{Theorem}

\newtheorem*{lemma*}{Lemma}

\allowdisplaybreaks
\begin{document}

\newcommand{\nop}{\circ} 
\newcommand{\yesp}{\star}
\newcommand{\rod}{\bullet}
\newcommand{\rodr}{\stackrel{\rightarrow}{\bullet}}
\newcommand{\rodl}{\stackrel{\leftarrow}{\bullet}}

%opening
\title{An exactly solvable model for RNA polymerase during the elongation stage}

\author{Ngo P.N. Ngoc\footnote{Institute of Research and Development, Duy 
Tan University, Da Nang 550000, Vietnam and Faculty of Natural Sciences, 
Duy Tan University, Da Nang 550000, Vietnam. email: 
ngopnguyenngoc@duytan.edu.vn} \and Vladimir Belitsky\footnote{Instituto de 
Matemática e Estátistica, Universidade de São Paulo, Rua do Matão, 1010, CEP 
05508-090, São Paulo - SP, Brazil. email: belitsky@ime.usp.br} \and Gunter M. 
Sch\"{u}tz\footnote{IAS 2, Forschungszentrum J\"ulich, 52425 J\"ulich, 
Germany. email: g.schuetz@fz-juelich.de}}
%\author{Vladimir Belitsky\thanks{Instituto de Matemática e Estátistica, Universidade de São Paulo, Rua do Matão, 1010, CEP 05508-090, São Paulo - SP, Brazil. \\email: belitsky@ime.usp.br}, Ngo P.N. Ngoc \thanks{Institute of Research and Development, Duy Tan University, Da Nang 550000, Vietnam.} \thanks{Faculty of Natural Sciences, Duy Tan University, Da Nang 550000, Vietnam. \\ email: ngopnguyenngoc@duytan.edu.vn}, Gunter M. Sch\"{u}tz\thanks{Departamento de Matemática, Instituto Superior Técnico, Universidade de Lisboa, Av. Rovisco Pais 1, 1049-001 Lisbon,Portugal. \\ email: gschuetz04@yahoo.com}}

\date{4/07/2024}

\maketitle

\begin{abstract}
We consider a Markovian model for the kinetics of RNA Polymerase (RNAP) 
which provides a physical explanation for the 
phenomenon of cooperative pushing during transcription elongation 
observed in biochemical experiments 
on Escherichia coli and yeast RNAP. To study how backtracking of RNAP
affects cooperative pushing we incorporate into this model backward 
(upstream) RNAP moves. With a rigorous 
mathematical treatment of the model we derive conditions on the mutual 
static and kinetic interactions between RNAP under which backtracking 
preserves cooperative pushing. This is achieved by exact computation 
of several key properties in the steady state of this model, including the 
distribution of headway between two RNAP along the DNA template and 
the average RNAP velocity and flux.
\end{abstract}


\section{Introduction}

RNA polymerase (RNAP) is an enzyme that functions as a molecular motor 
responsible for transcribing the genetic information encoded in the DNA base 
pair sequence into RNA \cite{Albe13}. The transcription process unfolds in 
three distinct phases: initiation, elongation, and termination. At initiation RNAP 
binds to a specific region on the DNA called promoter sequence and starts 
separating the two strands of the DNA, thus creating the conditions for the 
onset of transcription elongation. Forming the transcription elongation 
complex (TEC) the RNAP proceeds along the DNA 
base pairs and within this complex, the enzyme polymerizes the monomeric 
subunits of
an RNA by adding nucleotides in accordance with the corresponding 
sequence on the DNA template, called transcription elongation. Each 
elongation step involves a catalytic mechanism encompassing 
several key stages, including as elaborated in \cite{Bai,Wang1998,Chow13} ,
(1) binding of nucleoside triphosphate (NTP), (2) 
hydrolysis of NTP, (3) release of pyrophosphate (PP$_i$) as a product of 
hydrolysis, and (4) concurrent forward movement of RNAP along the DNA 
template by one base pair, called translocation. 
Termination marks the conclusion of the transcription elongation process, 
occurring when the TEC encounters a specific termination sequence
and the RNAP detaches from the template DNA.

A model for the kinetics of translocation must take into account that
thermal noise and other factors such as sufficient supply of NTP introduce
randomness into the amount of time that is necessary to complete the
mechanochemical cycle involved in a translocation step. Also, many RNAP 
move simultaneously on the same promoter sequence, 
so that one cannot ignore their mutual interactions. In particular, due to steric 
hindrance they cannot occupy the same region on the DNA template which is 
incorporated  in most modeling approaches to molecular motor traffic as a 
hard core repulsion 
\cite{Scha10,Lipo10,Chou11,Kolo15,Graf17,Fang19,Cava23}. It is remarkable 
that with this steric excluded volume interaction alone, one can successfully 
capture the appearance of RNAP ``traffic jams'' which is a collective 
phenomenon that occurs when a pausing RNAP prevents a trailing RNAP 
from moving forward and thus leads to a reduction of the average flux of 
RNAP along the DNA and consequently decreases the rate of elongation. 
Using the asymmetric simple exclusion process (ASEP) 
\cite{Derr98,Schutz2001,Liggett2010} as a prototypical model for
molecular motors \cite{Scha10} this has been demonstrated in the context of
protein synthesis by ribosomes in \cite{MacD68,Schu97} and more rigorously
and in considerable detail from a mathematical perspective in 
\cite{Ferr91,Derr93,Ferr94,Dudz00,Beli02,deGi06}.

However, as empirically demonstrated both in vitro and in vivo already some 
time ago for Escherichia coli \cite{Epshtein2003-1,Epshtein2003-2,Jin10} and 
for yeast \cite{Saek09}, interactions between RNAP may also be cooperative 
and lead to an enhancement of the rate of elongation. This has been argued 
to originate in a process where trailing RNAP "pushes" the leading RNAP out 
of pause sites \cite{Galb11,Stev20,Wang24}. To account for this mechanism 
in stochastic models of transscription elongation, RNAP interactions more 
complex than just excluded volume due to steric hindrance need to be 
considered.

Belitsky and Sch\"{u}tz in \cite{Belitsky2019-1,Belitsky2019-2} introduced
such a model of RNAP interactions which predicts conditions under 
which either jamming or cooperative pushing arise. This model is a 
generalization of the asymmetric simple exclusion process (ASEP) 
\cite{Derr98,Schutz2001,Liggett2010}, originally introduced in the seminal 
work \cite{MacD68} on the kinetics of protein synthesis by ribosomes and 
since then widely used as a starting point for modelling many different kinds 
of molecular motors \cite{Scha10}. The generalized ASEP first 
introduced in \cite{Belitsky2019-1} augments the original ASEP with a 
next-nearest interaction and with an internal degree of freedom 
to account in the spirit of \cite{Tripathi2008} for the mechanochemical cycle 
that RNAP undergoes during transcription elongation. The transition rates 
for translocation in this model depend on
the presence of nearby RNAP to account for both blocking and pushing. 

It should be stressed that on the microscopic level of interactions between 
individual RNAP these configuration-dependent rates describe the mutual 
interactions of blocking and pushing between neighboring RNAPs in an 
explicit way detailed below, while on the macroscopic empirical level studied 
in the biochemical experiments
\cite{Epshtein2003-1,Epshtein2003-2,Jin10,Saek09}, 
those rates lead in an intricate way to the {\it collective} phenomena jamming 
and cooperative pushing. Shedding light on the emergence of these 
phenomena requires a detailed mathematical analysis of the macroscopic 
properties of the microscopic model that is difficult to achieve by commonly 
used numerical simulations or analytic approximation schemes such as mean 
field theories. 

Such a mathematical analysis is possible for the model of 
\cite{Belitsky2019-1} which is an exclusion process with short-range 
interactions in addition to pure excluded volume interactoin. Studying the
stationary distribution has revealed that an enhancement
of the rate of elongation cannot be explained by
the mere existence of microscopic pushing between RNAP. For cooperative 
pushing in macroscopic to emerge, 
sufficiently strong repulsive interactions in addition to excluded volume
interaction which are reflected in pushing above a certain
critical strength are necessary.
However, given the complexity of the mechanochemical translocation step
that is only very partially taken into account in \cite{Belitsky2019-1}
this deeper insight into cooperative pushing is not yet fully satisfactory. 
The robustness of the argument for a minimal critical pushing strength 
and how pushing competes with potentially opposing forces still need to 
be probed.

Indeed, a drawback of the model of \cite{Belitsky2019-1} is the feature of 
purely unidirectional motion of RNAP along the DNA template. This 
simplification does not capture backtracking, i.e., a backward jumps of the 
RNAP during transcription \cite{Shae03,Galb07}. This may happen 
when the polymerase tries to incorporate a noncognate NTP and plays a role 
in error correction to enhance transcription fidelity
\cite{Saho13}. Despite a generically small error rate \cite{Gout13,Gout17} error
correction is an important process since a single mutated RNA transcript
can have a large effect. Since backtracking is an upstream movement
of RNAP against the mean forward (downstream) flow due to translocation, 
one might wonder whether backtracking would not only reduce the rate of 
RNA polymerization but also overcome the effect of pushing between
individual RNAP and thus prevent the emergence cooperative pushing, 
i.e., the boosting of the efficiency of transcription by pushing of stalled 
RNAP.

Backtracking has been investigated not only as error correction mechanism 
but in some detail in \cite{Galb11} on a molecular level, showing that 
translocation of RNAP might occur through a power stroke. A recent study 
based on a different exclusion process with configuration dependent rates 
investigated the interplay of pushing and backtracking \cite{Zuo22}.
The microscopic dynamics of that model, which was studied analytically 
using a mean-field approximation, bias the system to more likely incorporate 
the noncognate nucleotide. Here we address the role of backtracking in 
cooperative pushing by extending the model of
\cite{Belitsky2019-1,Belitsky2019-2} to include backward translocation
in a way that maintains the rigorous mathematical tractability.
As an advantage of this approach we note that  one can still calculate {\it 
exact}  stationary bulk properties
of the kinetics of transcription elongation of this more sophisticated exclusion 
model, without any uncontrolled 
approximation like mean field theory and thus explore 
 {\it quantitatively} how
macroscopic stationary properties arise from kinetic
interactions between single RNAP that are encoded in the microscopic 
transition rates.

In particular, one can compute how in the presence of backtracking the flux 
of RNAP along the DNA template depends on the density of RNAP and the 
strength of interactions between them, which is the purpose of the present 
work. The focus is on the important role of collisions between RNAP during 
transscription elongation, reviewed recently in \cite{Wang24}.
It is not intended to provide a comprehensive analysis of all 
microscopic mechanisms of backtracking and translocation that
occur during transcription elongation but to highlight the role of the
backtracking that arises from the reverse 
mechanochemical process
associated with the main process of translocation. 
The regulation of the RNAP
density by the kinetics of initiation, termination \cite{MacD68,Schu93b,Derr93a}, and bulk factors 
such as
bulk attachment and detachment of RNAP 
\cite{Schu95,Parm03,Popk03,Gome19}, or defects 
\cite{Chai09,Thol12,Appe15,Erdm20,Schu21} is not considered. We also
neglect interactions between RNAP that may arise from DNA supercoiling 
\cite{Chat21}. Such interactions would lead to long-range interactions along the
template which is out of the scope of the present framework. We only mention 
that exclusion processes with nontrivial exactly solvable stationary 
distributions and 
long-range interactions may be constructed along the lines described in 
\cite{Kare17}.
Notice that although we treat a particular backtracking mechanism
separately, we aim, in future work, at a unified treatment of such models, via 
the exact mathematical consistency approach used in
\cite{Belitsky2019-1,Belitsky2019-2} and in the present paper.

To facilitate the distinction between microscopic processes involving
interactions between individual RMAP and the collective macroscopic result 
of these
interactions that can be observed in biochemical experiments e.g. in terms
of the rate of elongation, and understand how the interplay of microscopic 
processes generate collective behaviour we shall from now on refer to {\it 
blocking} and {\it 
pushing} when refering to microscopic forces acting on RNAP and to {\it 
jamming}
and {\it boosting} when discussing the resulting collective phenomena of
a reduction or enhancement respectively of the flux of RNAP along the
DNA template and thus to a corresponding change in the rate of elongation.

 The paper is organized as follows. In the following section, we present the 
 mathematical setting of our model. We begin by defining the state space of 
 RNAP configurations allowed in the framework of the model and an exposition 
 of the microscopic model dynamics (Subsection \ref{modelstatespace}). In 
 Subsection 
 \ref{modelstatinarydistribution} we introduce the stationary distribution of the 
 model and present the central mathematical result of this work. In the third 
 section, we explore various stationary properties of the model, including the 
 RNAP headway distribution, average excess, and the average elongation rate 
 in terms of the stationary RNAP flux and discuss the impact of backtracking 
 on these quantities. In the concluding section, we provide a concise summary 
 of our findings and present some open problems that are triggered by the 
 present results.

\section{Methods}
%\section{The model for interacting RNAP in the elongation stage of the 
%RNA transcription process}
\label{Sec:Modeldefinition}

The basic idea of our consistency approach is to introduce a generalized Ising 
measure in a parametric form and determine transition rates such that this 
measure is stationary. The transitions that occur with these rates are chosen 
to mimick the translocation process of RNAP. This approach involves a series 
of steps. (i) We envisage the DNA template as a one-dimensional lattice with a 
length of $L$, where individual lattice sites are numbered from $1$ to $L$. 
RNAPs are depicted as rods covering $l_{rod}$ consecutive sites, reflecting 
the physical reality where each RNAP covers $l_{rod}$ nucleotides 
\cite{MacD68,Laka03,Shaw03,Scho04,Gupt11,Duc18}. (ii) We propose a 
stationary distribution with {\it static interactions} as in \cite{Belitsky2019-1} 
that take into account {\it static} interactions between these rods, viz., 
excluded volume interaction like in the ASEP and a short-range interaction 
with the nearest RNAP on the lattice, leading to phenomena beyond what the 
Simple Exclusion Process can demonstrate. Weak logarithmic long-range 
interactions of entropic origin \cite{Carl02,Bar09,Hirs11,Beli24}
are neglected.
(iii) Following \cite{Tripathi2008} we define the chemical cycle that an RNAP undergoes in each translocation step in a reduced fashion in terms of transition rates between two states in which each RNAP may exist. (iv) We postulate the transition rates governing the motion of the RNAP along the DNA template which describe {\it kinetic} interactions that reflect the static excluded volume and nearest neighbor interactions.
For these we derive a consistency condition that ensures that the envisioned 
distribution is indeed stationary for the dynamics specified by those rates. 

\subsection{Mathematical modelling of the process}\label{modelstatespace}

As mentioned above, we represent the DNA template as a one-dimensional 
lattice with a length $L$ in units of a step length of $\delta \approx 0.34$ nm
determined by the size of single base pair. 
In our approach, we do not differentiate between 
RNAP and TEC, even in the presence of the intricate TEC structure. Instead, 
we simplify the TEC by modeling it as a hard rod with a defined length denoted 
as $l_{rod}$. This parameter represents the extent of nucleotides covered 
by an RNAP. In a scenario with $N$ RNAPs on the lattice, they are 
consecutively labeled by integers, with $i$ ranging from 1 to $N$. 
Specifying the position $k_i$ of an RNAP on the lattice then only 
requires knowledge of the position of the leftmost nucleotide it covers, which 
we refer to the position of the RNAP. Due to 
excluded volume interaction, no lattice site can be 
simultaneously covered by more than one RNAP. Furthermore, to 
account for the 
mechanochemical cycle we allow for RNAP to occur in two distinct 
polymerization states: one without PP$_i$ bound (state 1) and the other with 
PP$_i$ bound (state 2). 

Once RNAP has released PP$_i$, it can advance along the DNA template by a 
single base pair, equivalent to a step length 1 on the lattice. In 
terms of our lattice model, this translocation thus implies that an RNAP positioned 
at location $k_i$ in state 1 can progress one site forward, shifting from $k_i$ 
to $k_i + 1$, provided that the site $k_{i} +l_{rod}$ is unoccupied. 
Conversely, 
the RNAP can move in reverse, leading to the depolymerization of RNA from 
its position at $k_i$ to $k_{i} - 1$, provided that the site $k_{i} - 1$ is vacant. 
This backtracking occurs only when the RNAP is in state 2, indicating that 
PP$_i$ is bound to it. Therefore, the presence and status of RNAPs along 
the same DNA segment can be described at any moment of time by their 
positions and states. Thermal noise, availability of NTP and other molecules
required for these processes to happen leads to a translocation dynamics
of individual RNAP that is subject to randomness.
To define the stochastic mathematical model for this random process we 
refer to rods rather than RNAP and to the lattice rather than DNA template. 

\subsubsection{Rod configurations}

\begin{figure}[h]
	\centering
	\includegraphics[width=0.33\textwidth]{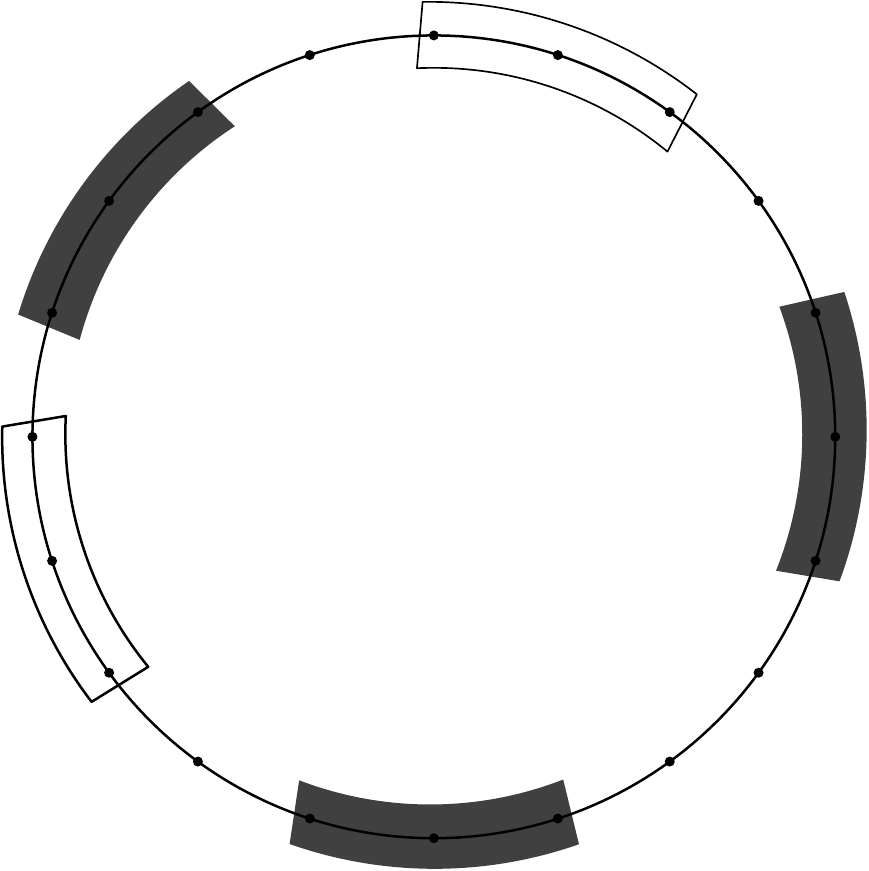}
	\caption{An allowed configuration on a ring with $L = 20, N = 5, l_{rod} = 
	3$. 
	Black rods are in state 1 and blank rods are in state 2.}
	\label{ring_rods}
\end{figure}

We define a complete configuration of rods on the lattice, denoted as 
$\boldsymbol{\eta}$, by using the set of positions $k_i \in \{1,2,\dots,L\}$ and 
corresponding states $\alpha_i \in\{1,2\}$ of the rods. 
It is important to note that if $k_i$ represents the position of a rod, then 
$k_i + l_{rod} - 1$ represents the lattice position of the "front" side of the rod. 
In an allowed configuration the ordering condition 
$k_{i+1} \geq k_i +l_{rod}$ must be satisfied due to the excluded volume rule.
 This constraint is expressed 
as $k_{i+1} \ge k_i + l_{rod}$ and we say that two rods $i$ and $i+1$ are 
neighbors 
when the front end of rod $i$ and the left edge of rod $i+1$ occupy 
neighboring lattice sites, i.e., when $k_{i+1} = k_i+l_{rod}$.
Since we are interested only in the elongation stage of transcription, 
we take a lattice of $L$ sites with periodic boundary conditions, see Fig. \ref{ring_rods}.

\subsubsection{Transition rates for the mechanochemical cycle}

\begin{figure}[h]
	\centering
	\includegraphics[width=0.4\textwidth]{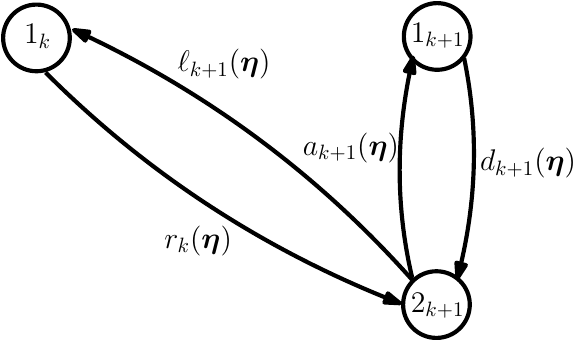}
	\caption{Minimal scheme of the mechano-chemical cycle of an RNAP. The 
	RNAP without PP$_i$ bounds to it is in state 1 and with PP$_i$ is in state 2. 
	The integer subscript $k$ labels the position of the RNAP on the DNA 
	template. Within this scheme, the $i^{th}$ RNAP, situated in state 1 at 
	position $k$ and denoted as $1_k$, possesses the ability to move from 
	base pair $k$ to $k + 1$. This translocation is contingent on the current 
	system configuration and is quantified by the configuration-dependent rate 
	$r_i(\boldsymbol{\eta})$. However, the subsequent translocation step for 
	the RNAP can only occur following the release of PP$_i$, a process 
	governed by a rate that is denoted as $a_i(\boldsymbol{\eta})$. This 
	transition leads the RNAP from state $2_{k + 1}$ to state $1_{ k + 1}$. In the 
	event that the RNAP is positioned at base pair $k+1$ and in state $2_{k+1}$, 
	it can move back to base pair $k$ through the depolymerization of RNA. 
	This backward movement is associated with a rate represented as 
	$\ell_i(\boldsymbol{\eta}$), resulting in a transition from state $2_{k+1}$ to 
	state $1_{k}$. Finally, the association of PP$_i$ is accompanied by a rate 
	$d_{i}(\boldsymbol{\eta})$, enabling the transition from state $1_{k+1}$ 
	to state $2_{k+1}$.
	}
	\label{fig_RNAP_d1}
\end{figure}

The rate at which the forward step, i.e., translocation, of rod $i$ occurs is 
denoted as $r_i(\boldsymbol{\eta})$. 
The rate of the backward movement. i.e., backtracking, is represented as 
$\ell_i(\boldsymbol{\eta})$. Additionally, we denote the rate of PP$_i$ release 
as $a_i(\boldsymbol{\eta})$ and the rate of PP$_i$ binding as 
$d_i(\boldsymbol{\eta})$. This minimal reaction scheme aligns with the 
description found in \cite{Wang1998,Belitsky2019-1,Tripathi2008} for a single 
RNAP and is 
illustrated in Fig. \ref{fig_RNAP_d1}.

\begin{figure}[H]
	\centering
	\subfloat[\centering $r_i(\boldsymbol{\eta}) = r$]{{\includegraphics[width=7.5cm]{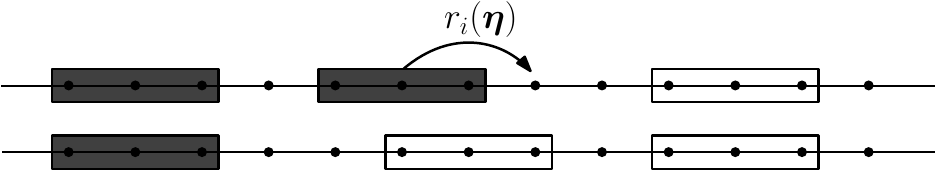} }}\label{Fig:rod1}
	\qquad
	\subfloat[\centering $r_i(\boldsymbol{\eta}) = r(1 + 
	r^{\yesp\rodr})$]{{\includegraphics[width=7.5cm]{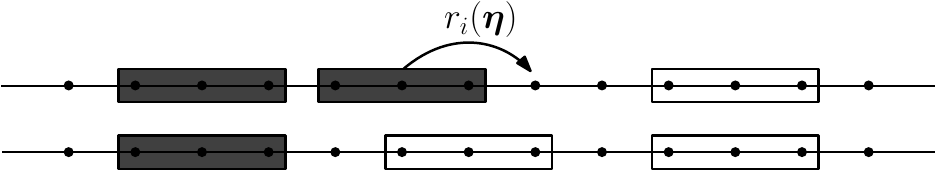} }}%
	\qquad
	\subfloat[\centering $r_i(\boldsymbol{\eta}) = r(1 + r^{\yesp\rodr}+ 
	r^{\rodr\nop\yesp})$]{{\includegraphics[width=7.5cm]{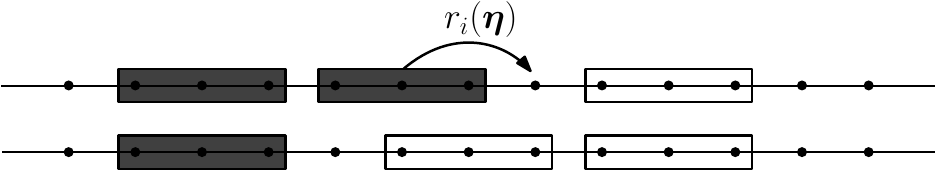} 
	}}%
	\qquad
	\subfloat[\centering $r_i(\boldsymbol{\eta}) = 0$]{{\includegraphics[width=7.5cm]{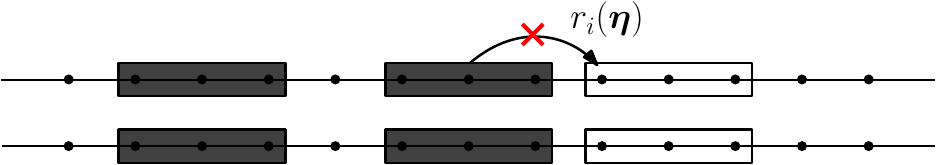} }}%
	\caption{Some translocation rates $r_i(\boldsymbol{\eta})$ for a rod in state 1. 
In these visual representations, black rods are depicted in state 1, while blank rods are shown in state 2. }%
	\label{jump_rate_00}
\end{figure}

\begin{figure}[H]
	\centering
	\subfloat[\centering $a_i(\boldsymbol{\eta}) = a(1+a^{\yesp\rod})$]{{\includegraphics[width=7.5cm]{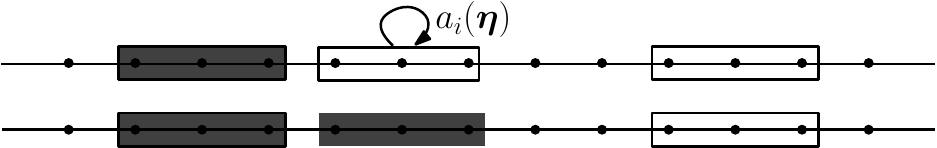} }}%
	\qquad
	\subfloat[\centering $d_i(\boldsymbol{\eta}) = d(1 + d^{\yesp\rod})$]{{\includegraphics[width=7.5cm]{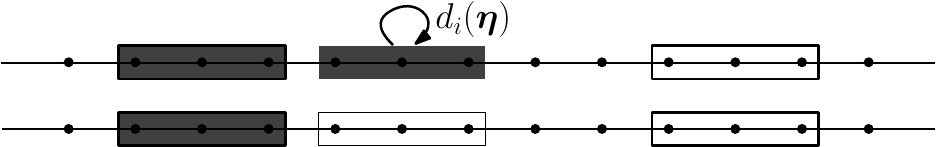} }}%
	\qquad
	\subfloat[\centering $a_i(\boldsymbol{\eta}) = a(1 + a^{\yesp\nop\rod}+ a^{\rod\nop\yesp})$]{{\includegraphics[width=7.5cm]{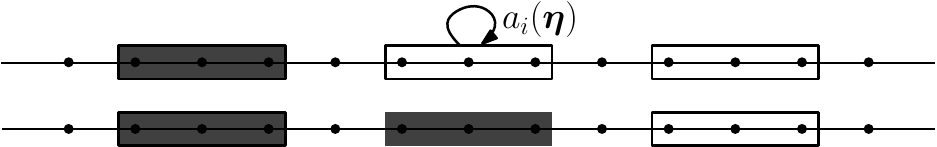} }}%
	\qquad
	\subfloat[\centering $d_i(\boldsymbol{\eta}) = d(1 + d^{\yesp\nop\rod}+ d^{\rod\nop\yesp})$]{{\includegraphics[width=7.5cm]{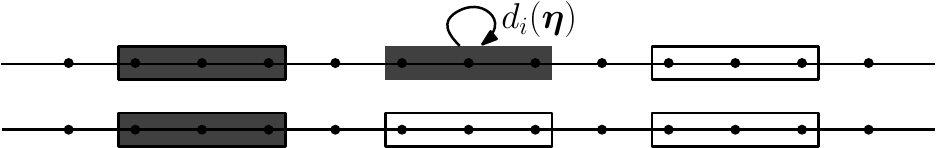} }}%
	\caption{Some binding and release rates $a_i(\boldsymbol{\eta})$ and $d_i(\boldsymbol{\eta})$. Black rods represent RNAP in state 1, while the blank rods signify RNAP in state 2 (with PP$_i$ bound to it).}%
	\label{release_rate_00}
\end{figure}

Let $\boldsymbol{\eta}$ be an allowed configuration with the coordinate 
vector $\mathbf{k} = (k_1,...,k_N)$ and state vector $\boldsymbol{\alpha} = 
(\alpha_1,...,\alpha_N)$. The above-mentioned 
rates are of the form 
\begin{align}
	r_i(\boldsymbol{\eta}) = & 
\  r \, \delta_{\alpha_i,1} 
(1+r^{\yesp\rodr}\delta_{k_{i-1}+ l_{rod},k_i} + 
r^{\rodr\nop\yesp}\delta_{k_{i}+ l_{rod}+1, 
	k_{i+1}})(1-\delta_{k_{i}+ l_{rod} ,k_{i+1}}),\label{rate1_d1} \\
	\ell_i(\boldsymbol{\eta}) = & \ \ell \, \delta_{\alpha_i,2} 
	(1+\ell^{\yesp\nop\rodl}\delta_{k_{i-1}+l_{rod} +1,k_i} + 
	\ell^{\rodl\yesp}\delta_{k_{i}+l_{rod} , k_{i+1}}) 
	(1-\delta_{k_{i-1}+l_{rod} ,k_{i}}), 
	\label{rate2_d1}\\
\begin{split}
a_i(\boldsymbol{\eta}) = 
& \ a \,\delta_{\alpha_i,2}  
[1+a^{\yesp\rod}\delta_{k_{i-1}+l_{rod},k_i} + 
a^{\rod\yesp}\delta_{k_{i}+l_{rod},k_{i+1}} + 
a^{\yesp\rod\yesp}\delta_{k_{i-1}+l_{rod},k_{i}}
\delta_{k_{i}+l_{rod},k_{i+1}} \\ 
& + a^{\yesp\nop\rod}(1-\delta_{k_{i-1}+l_{rod},k_{i}})
\delta_{k_{i-1}+l_{rod}+1,k_{i}}
+ a^{\rod\nop\yesp}(1-\delta_{k_{i}+l_{rod},k_{i+1}})
\delta_{k_{i}+l_{rod}+1,k_{i+1}}],
	\end{split}\label{rate3_d1}\\
	\begin{split}
		d_i(\boldsymbol{\eta}) = & \ d \,\delta_{\alpha_i,1}  	
		[1+d^{\yesp\rod}\delta_{k_{i-1}+l_{rod},k_i} + 
		d^{\rod\yesp}\delta_{k_{i}+l_{rod},k_{i+1}} + 
		d^{\yesp\rod\yesp}\delta_{k_{i-1}+l_{rod},k_{i}}\delta_{k_{i}+l_{rod},k_{i+1}}
		 \\ 
		& +  
		d^{\yesp\nop\rod}(1-\delta_{k_{i-1}+l_{rod},k_{i}})\delta_{k_{i-1}+l_{rod}+1,k_{i}}
		 			+ 
		d^{\rod\nop\yesp}(1-\delta_{k_{i}+l_{rod},k_{i+1}})\delta_{k_{i}+l_{rod}+1,k_{i+1}}].
	\end{split}\label{rate4_d1}
\end{align}
In this setting, the transitions are 
contingent on the configuration as given by the Kronecker-$\delta$ factors, and their
rates depend on 16 parameters all of which 
describe the kinetic interactions between neighboring RNAPs.
The notation for the rates and kinetic interaction parameters is chosen as follows.
\begin{itemize}
\item[-] The subscript $i$ on the rates refers to the rod with label $i$ at 
position $k_i$ in state $\alpha_i$ in the configuration $\boldsymbol{\eta}$.
\item[-] The parameters $r,\ell,a,d$ are rates in units of seconds. They would 
be 
the transition rates of the rods if 
only excluded volume interaction was taken into account. Hence we call them 
{\it bare rates}. 
\item[-] The parameters with superscripts are dimensionless numbers that 
describe the kinetic interactions by multiplying the 
bare rates in a way that depends on the location $k_{i\pm 1}$ of the 
neighboring rods $i\pm1$ as determined by the Kronecker-$\delta$ factors.  
We call these quantities 
{\it kinetic interaction parameters}.
\item[-] The quantities denoted by $r, r^{\yesp\rodr}, r^{\rodr\nop\yesp}$ 
determine the jump rates to the right (translocation).
\item[-] The quantities denoted by $\ell, \ell^{\yesp\nop\rodl}, 
\ell^{\rodl\yesp}$ determine the jump rates to the left (backtracking).
\item[-] The quantities denoted by $a,a^{\yesp\rod}, a^{\rod\yesp}, 
a^{\yesp\rod\yesp}, 
a^{\yesp\nop\rod}, a^{\rod\nop\yesp}$ determine the release rates of PP$_i$.
\item[-] The quantities denoted by $d,d^{\yesp\rod}, d^{\rod\yesp}, 
d^{\yesp\rod\yesp}, d^{\yesp\nop\rod}, d^{\rod\nop\yesp}$ determine the binding rates of PP$_i$.
\item[-] The superscript $\rod$ refers to rod $i$ and an arrow above it
indicates the jump direction.
\item[-] The superscript $\yesp$ refers to a neighboring rod $i\pm 1$ and
is placed to the right of $\rod$ for kinetic interactions with the right neighboring rod
$i+1$ and to the left of $\rod$ for kinetic interactions with the left neighboring rod
$i-1$ or on both sides for kinetic interactions influenced by both neighbors.
\item[-] The superscript $\nop$ refers to one empty site next to rod $i$ and is 
placed to the right (left) of $\rod$ for one empty site to the right (left) of rod 
$i$.
\item[-] In the absence of the superscript $\nop$ the next rod $i\pm 1$ 
indicated by $\yesp$
in the superscript to the right or left of $\rod$
is on a nearest neighbor site (without an empty site in between) while the presence 
the superscript $\nop$ there is one empty site between rod $i$ and rod $i\pm 1$,
and we say that rod $i\pm 1$ is on a next-nearest neighbor site.
\end{itemize}
In Figs. \ref{jump_rate_00} and \ref{release_rate_00} some of these transitions
are illustrated, with Figs. \ref{jump_rate_00}(a)-(d)
showing how the rate of translocation depends on the presence rods on neighboring sites ,
and Figs. \ref{release_rate_00}(a)-(d) displaying various
transitions between states 1 and 2.


The excluded volume interaction is taken into account by the
overall factors $(1-\delta_{k_{i}+l_{rod},k_{i+1}})$ and 
$(1-\delta_{k_{i-1}+l_{rod},k_{i}})$ in the rates for translocation and
backtracking which forbid jumps onto an occupied site, corresponding to 
blocking. Below we indicate this by definining hypothetical interaction
parameters
$r^{\rodr\yesp} = \ell^{\yesp\rodl} = -1$ for jumps onto occupied
sites that would violate the exclusion rule.
The factors
$\delta_{k_{i-1}+l_{rod}+1,k_{i}}$ capture the kinetic next-nearest neighbor
interaction.
The overall factors
$\delta_{\alpha_i,\beta}$ ensure that the transitions between the chemical 
states 1 and 2 occur as described by the simplified mechanochemical cycle
we consider in this work.

\subsubsection{Choice of rates and kinetic interaction range}

\paragraph{Bare rates:}
In the setting of Wang \textit{et al.} \cite{Wang1998} the bare rates $a, r, \ell$, and $d$ take the values
\begin{equation}\label{parameters_d1}
	r =\text{[NTP]}(\mu M)^{-1}s^{-1},\quad \ell = 0.21s^{-1},\quad a = 31.4s^{-1},\quad  d = \text{[PP}_i\text{]}(\mu M)^{-1}s^{-1}.
\end{equation}
Here [NTP] and [PP$_i$] are the NTP and PP$_i$ concentrations which 
following \cite{Tripathi2008} 
are chosen as $[NTP] = 10^{-3}$, $[PP_i] 
= 10^{-5}$. We stress that is not the purpose of this study to predict 
elongation rates for any concrete biological process but to study how 
interactions between RNAP affect the elongation rate qualitatively.
Hence we adopt these specific empirical parameters as reference constants
throughout this work.

\paragraph{Kinetic interaction parameters:} We have no empirical data on the 
kinetic interaction 
parameters at our disposal. Hence they are taken as variables and the main
characteristics of the model are computed for different values of these
variables to explore how the main quantitites depend on these unknown quantities
which, in principle, are measurable in experiments.
To ensure the positivity of the rates and ergodicity of the 
process
for all allowed configurations, all interaction rates must be individually larger than or equal to -1 and in combination with others satisy the inequalities 
$r^{\yesp\rodr} + 
r^{\rodr\nop\yesp}\geq -1$, $\ell^{\yesp\nop\rodl} + \ell^{\rodl\yesp} \geq -1$, 
$a^{\yesp\rod} + a^{\rod\yesp} \geq -1$, $a^{\yesp\rod} + a^{\rod\nop\yesp} 
\geq -1$, $a^{\yesp\nop\rod} + a^{\rod\yesp} \geq -1$, $a^{\yesp\nop\rod} + 
a^{\rod\nop\yesp} \geq -1$, $a^{\yesp\rod\yesp}\geq -1$. 
Depending on the sign of the interaction parameters for 
translocation
they describe kinetic repulsion or kinetic attraction as follows.

When $r^{\yesp\rodr}>0$ the bare translocation rate $r$ is 
increased
by the presence of a trailing RNAP which means RNAP pushing to the right.
Similarly, $\ell^{\rodl\yesp} >0$ increases the bare backtracking rate $\ell$ in 
the presence of a neighboring RNAP upstream, which 
means RNAP pushing to the left. As mentioned in the introduction, we refer 
these processes, which correspond to a kinetic repulsion, as pushing, as 
opposed to the boosting (i.e., cooperative 
pushing) reported for the biochemical
experiments \cite{Epshtein2003-1,Epshtein2003-2,Saek09,Jin10}. We stress
once more that as shown in \cite{Belitsky2019-1}, RNAP pushing on the level 
of individual RNAP does {\it not} automatically imply boosting.

When $r^{\rodr\nop\yesp}<0$ the bare translocation rate $r$ is 
reduced
by the presence of a next-nearest neighbor upstream RNAP. We refer to
this effect as blocking enhancement as it corresponds to a kinetic repulsion
that is of longer interaction range, but less strong than the full suppression of 
translocation in the presence nearest neighbor upstream RNAP due to
excluded volume interaction. Similarly, $\ell^{\yesp\nop\rodl}<0$ corresponds
to repulsive blocking enhancement for backtracking. We recall that
blocking does not necessarily imply jamming.

On the contrary, when $r^{\yesp\rodr} < 0$ then the bare 
translocation rate 
$r$ is reduced by the presence of a trailing nearest neighbor RNAP, and 
similarly when $\ell^{\rodl\yesp} < 0$ then the bare 
backtracking rate $\ell$ decreases due to the presence of a nearest neighbor 
upstream RNAP. These effects may be described as ``clinging'', 
corresponding
a kinetic attraction. Also $r^{\rodr\nop\yesp}>0$ and 
$\ell^{\yesp\nop\rodl}>0$ 
describe a form of kinetic attraction due to a ``pulling'' by a next-nearest 
neighbor RNAP upstream in case of translocation or downstream in case
of backtracking.

\paragraph{Interaction range:} 
Notice that for $r^{\rodr\nop\yesp} = \ell^{\yesp\nop\rodl} = 0$
the transition rates depend only on whether the {\it nearest} neighbor site
if rod $i$ is occupied by another rod via excluded volume interaction
and the through the interaction parameters $r^{\yesp\rodr},\ell^{\rodl\yesp}$, 
as opposed to $r^{\rodr\nop\yesp} \neq 
0$
or $ \ell^{\yesp\nop\rodl} \neq 0$ when the transition rates depend also on
occupation of the next-nearest neighbor site. We call the simplified scenario
$r^{\rodr\nop\yesp} = \ell^{\yesp\nop\rodl} = 0$ {\it minimal interaction 
range} while otherwise speak of {\it extended interaction 
range}. 

We summarize the role of the interaction terms in itemized form.
\begin{align}
%\bullet \mbox{ Static repulsion:} \quad &  y>1 \\
\bullet \mbox{ Kinetic repulsion: } &   \left\{ \begin{array}{ll}
r^{\yesp\rodr} = \ell^{\rodl\yesp} = -1 & \quad \mbox{(Blocking)} \\
r^{\rodr\nop\yesp}<0, \, \ell^{\yesp\nop\rodl}<0 & \quad \mbox{(Blocking 
enhancement)} \\
r^{\yesp\rodr}>0, \, \ell^{\rodl\yesp}>0 & \quad \mbox{(Pushing)} \hspace*{40mm} \\
\end{array} \right. \\[4mm]
%\bullet \mbox{ Static attraction: } &  y<1 \\
 \bullet \mbox{ Kinetic attraction:} &  \left\{ \begin{array}{ll}
r^{\yesp\rodr}<0, \, \ell^{\rodl\yesp}<0 & \quad \mbox{(Clinging)} \\
r^{\rodr\nop\yesp}>0, \, \ell^{\yesp\nop\rodl}>0 & \quad \mbox{(Pulling)} \\
\end{array} \right.
\end{align}
Kinetic repulsion allows for repulsive forces that
reach further than the on-site steric excluded volume
interaction implemented by taking $r^{\yesp\rodr} = \ell^{\rodl\yesp} = -1$. 
Kinetic attraction represents a stylized form of 
Lennard-Jones forces which are repulsive at very close distance (blocking due to the steric excluded volume
interaction), attractive at small distance (clinging and
pulling), and eventually absent at larger distances.
By introducing the notions of clinging and pulling we do not presume that these
mechanisms exist in any specific process of transcription elongation.
They are features that arise naturally in the RNAP model 
studied here and they may or may not have counterparts in biological systems. 
When all interaction parameters are taken to zero then only excluded
volume interaction is taken into account.

\subsubsection{Master equation}

In a nutshell, the Markovian microscopic dynamics unfold as follows. Each rod 
is associated with four random Poissonian clocks, labeled as 1, 2, 3, and 4, 
each operating with configuration-dependent rates denoted as 
$r_i(\boldsymbol{\eta}), d_i(\boldsymbol{\eta}), \ell_i(\boldsymbol{\eta})$ and 
$a_i(\boldsymbol{\eta})$, respectively. When one of these four clocks for rod 
$i$ activates, the following scenarios can occur:
\begin{itemize}
	\item  For a rod $i$ in state 1:
	\begin{itemize}
		\item If the clock is 3 or 4, no action takes place.
		\item If the clock is 1, the rod $i$ advances one site, provided that the 
		target site $k_i+ l_{rod}$ is unoccupied. Consequently, the coordinate of 
		the 
		$i^{th}$ rod changes to $k_i+1$, and its state instantly switches to 2.
		\item If the clock is 2, the position of the rod remains unchanged, but its state transitions to 2.
	\end{itemize}
	\item For a rod $i$ in state 2:
	\begin{itemize}
		\item If the clock is 1 or 2, there is no effect.
		\item If the clock is 3, the rod $i$ moves backward by one site, contingent 
		on the target site $k_i-1$ being unoccupied. Consequently, the 
		coordinate of the $i^{th}$ rod becomes $k_i-1$, and its state promptly 
		shifts to 1.
		\item If the clock is 4, the position of the rod remains unchanged, but its state changes to 1.
	\end{itemize}
\end{itemize}

With this definition
the master equation for the probability $\mathbb{P}_t(\boldsymbol{\eta})$ of finding the rods at time $t$ in the configuration $\boldsymbol{\eta}$ is as follows
\begin{align}\label{mastereq_d1}
	\dfrac{d}{dt}\mathbb{P}(\boldsymbol{\eta},t) 	 = & \ \sum_{i=1}^{N}\bigg[r_i(\boldsymbol{\eta}_{tlf}^i)\mathbb{P}(\boldsymbol{\eta}_{tlf}^i,t) + \ell_i(\boldsymbol{\eta}_{tlb}^i)\mathbb{P}(\boldsymbol{\eta}_{tlb}^i,t) +  a_i(\boldsymbol{\eta}_{rel}^i)\mathbb{P}(\boldsymbol{\eta}_{rel}^i,t) \nonumber\\
	& +  d_i(\boldsymbol{\eta}_{bin}^i)\mathbb{P}(\boldsymbol{\eta}_{bin}^i,t) - (r_i(\boldsymbol{\eta}) + \ell_i(\boldsymbol{\eta})+a_i(\boldsymbol{\eta}) + d_i(\boldsymbol{\eta}))\mathbb{P}(\boldsymbol{\eta},t)\bigg]
\end{align}
where $\boldsymbol{\eta}_{tlf}^i$ is the configuration that leads to 
$\boldsymbol{\eta}$ before a forward translocation of RNAP $i$ (i.e., with 
coordinate $k_i^{tlf} = k_i-1$ and state $\alpha_i^{tlf} = 3- \alpha_i$),  
$\boldsymbol{\eta}_{tlb}^i$  is the configuration that leads to 
$\boldsymbol{\eta}$ before a backward translocation of RNAP $i$ (i.e., 
$k_i^{tlb} = k_i+1, \alpha_i^{tlb} = 3-\alpha_i$), 
$\boldsymbol{\eta}_{rel}^i$ is the configuration $\boldsymbol{\eta}$ before 
PP$_i$ release at RNAP $i$ (i.e., $k_i^{rel} = k_i$ and  $\alpha_i^{rel} = 3 - 
\alpha_i$), 
and $\boldsymbol{\eta}_{bin}^i$ is the configuration leads to 
$\boldsymbol{\eta}$ before PP$_i$ binding at RNAP $i$ (i.e., $k_i^{bin} = k_i, 
\alpha_i^{bin} = 3-\alpha_i$). Notice here that due to periodicity, the positions 
$k_i$ of 
the rods are counted modulo $L$ and labels $i$ are counted modulo $N$.
The stationary master equation, denoted below by 
$\hat{\pi}(\boldsymbol{\eta})$, satisfies 
\eqref{mastereq_d1} with the left hand side taken to be zero.

\subsection{Stationary distribution}\label{modelstatinarydistribution}

Following \cite{Belitsky2019-1, Belitsky2019-2} the stationary
probability for the presence of rods at positions 
$\textbf{k} = (k_1,...,k_N)$ with states $\boldsymbol{\alpha} = 
(\alpha_1,...,\alpha_N)$ 
within the configuration $\boldsymbol{\eta}$ is expressed as follows:
\begin{equation}\label{invarmeas_d1}
	\hat{\pi}(\boldsymbol{\eta}) = \dfrac{1}{Z}\pi(\boldsymbol{\eta})
\end{equation}
where $\pi(\boldsymbol{\eta})$ is the Boltzmann weight which is of the form 
\begin{equation}\label{boltzmann_d1}
	\pi(\boldsymbol{\eta}) = \exp\left[-\dfrac{1}{k_BT}(U(\textbf{k}) + \lambda 
	B(\boldsymbol{\alpha})) \right].
\end{equation}
The quantity $T$ is an effective temperature that is considered as a constant.
The quantity $U(\mathbf{k})$ is the static short-range interaction energy described by
\begin{equation}\label{shortrangeintraction_d1}
	U(\mathbf{k}) = J\sum_{i=1}^{N}\delta_{k_{i+1}, k_i+l_{rod}}^L.
\end{equation}
A positive value of $J$ corresponds to repulsive static interaction between 
neighboring rods. 
Here, $\delta^L$ represents the Kronecker symbol, computed modulo $L$ 
due to the presence of periodic boundary conditions. 

The quantity
\begin{equation}
B(\boldsymbol{\alpha}) := \sum_{i=1}^{N}(3-2\alpha_i) = 
N^1(\boldsymbol{\eta}) - N^2(\boldsymbol{\eta})
\label{Bdef}
\end{equation} 
signifies the excess in the number $N^\alpha(\boldsymbol{\eta})$ of rods in 
state $i \in {1,2}$ in a configuration $\boldsymbol{\eta}$. The chemical 
potential $\lambda$ acts as a Lagrange multiplier, which parametrizes the 
mean excess and describes the fluctuations of the excess that arises from
the interplay of NTP hydrolysis and PP$_i$ release.
The partition function
\begin{equation}\label{partfunc_d1}
	Z = \sum_{\boldsymbol{\eta}}\pi(\boldsymbol{\eta})
\end{equation}
is not needed in explicit form in the computations below.
For the convenience of computation, one introduces
\begin{equation}\label{xydef}
	x= e^{\frac{2\lambda}{k_B T}},\ \ 	y= e^{\frac{J}{k_B T}},
\end{equation}
so that $x>1$ corresponds to an excess of RNAP in state 1 and repulsive static interaction corresponds to $y>1$. 

To ensure that the process governed by the dynamics 
\eqref{rate1_d1}--\eqref{rate4_d1} admits a measure of the form 
\eqref{invarmeas_d1} to be its invariant distribution, a price to pay is that 
the parameters of the model must satisfy the three consistency conditions
\begin{eqnarray}
x & = & \dfrac{r + d}{\ell + a} \label{stationarycondition1_d1_1} \\
y & = & \dfrac{1+r^{\yesp\rodr}}{1+r^{\rodr\nop\yesp}} = 
	\dfrac{1+\ell^{\rodl\yesp}}{1+\ell^{\yesp\nop\rodl}} 
	\label{stationarycondition1_d1_2} 
\end{eqnarray}
relating the parameters of the stationary distribution to the four bare rates and 
the four interaction parameters for translocation and backtracking
and the five consistency conditions
\begin{align}
	& xa a^{\yesp\rod} - d d^{\yesp\rod}= \dfrac{1}{1+x}(- r + x\ell ) 
	-\dfrac{x}{1+x}(- r r^{\yesp\rodr} + x\ell \ell^{\rodl\yesp} )\\
	& xa a^{\rod\yesp} - d d^{\rod\yesp}=  \dfrac{x}{1+x}(- r + x\ell ) 
	-\dfrac{1}{1+x}(- r r^{\yesp\rodr} + x\ell \ell^{\rodl\yesp} )\\
	& xa a^{\yesp\rod\yesp} - d d^{\yesp\rod\yesp}= -r r^{\yesp\rodr} + x\ell 
	\ell^{\rodl\yesp} \\
	& xa a^{\yesp\nop\rod} - d d^{\yesp\nop\rod}= \dfrac{1}{1+x}(r r^{\rodr\nop 
	\yesp} - x\ell \ell^{\yesp\nop\rodl} )  \\
	&  xa a^{\rod\nop\yesp} - d d^{\rod\nop\yesp}= \dfrac{x}{1+x}(r r^{\rodr
	\nop \yesp} - x\ell \ell^{\yesp\nop\rodl} ).
	\label{end_d1_1}
\end{align}
involving the nine interaction parameters for binding and release of PP$_i$.

This is proved rigorously in Appendix \ref{condtions_existence} and allows us 
to
present the main mathematical result of the present work as a formal theorem.

\begin{theorem}\label{main_theorem} If the parameters appearing in the 
rates \eqref{rate1_d1}--\eqref{rate4_d1} satisfy the consistency conditions
\eqref{stationarycondition1_d1_1}--\eqref{end_d1_1}, then the invariant 
measure of the rod process defined by the master equation
\eqref{mastereq_d1} is given by
	\begin{equation}\label{main_theorem_d1}
		\hat{\pi}(\boldsymbol{\eta}) = \dfrac{1}{Z}\left(\dfrac{r + d}{\ell + 
		a}\right)^{\sum_{i=1}^{N}-3/2+ \alpha_i}\left( \dfrac{1+r^{\yesp 
		\rodr}}{1+r^{\rodr\nop\yesp}} \right)^{-\sum_{i=1}^{N}\delta^L_{k_{i+1}, 
		k_i+l_{rod}}}
	\end{equation}
	where $Z$ is the partition function.
\end{theorem}

Notice that the excess part of the stationary distribution involving the
Langrange multiplyer $\lambda$ depends only on the bare transition
rates while the static interaction part of the stationary distribution involving the
Kronecker-$\delta$ terms depends only on the kinetic interaction parameters
which satisfy the symmetry \eqref{stationarycondition1_d1_2}. Remarkably, 
comparing this consistency condition with the role of the
interaction parameters shows that repulsive {\it kinetic}
interactions are consistent only with repulsive {\it static} interaction and 
similarly, attractive {\it kinetic}
interactions are consistent only with attractive {\it static} interaction.
While this is what one may expect on physical grounds the consistency 
conditions 
\eqref{stationarycondition1_d1_1} - \eqref{end_d1_1} between static and 
kinetic interaction
parameters are
{\it not} a feature built into the definition of the model but a purely
mathematical result
that comes out in the proof of the Theorem. 
The significance of the relations
between the parameters of the stationary distribution and the transition rates 
are discussed in the following 
section.

%Notice that if we set $d = \ell = 0$, the stationary conditions 
%\eqref{stationarycondition1_d1_1}--\eqref{end_d1_1} recover the ones in the 
%paper \cite{Belitsky2019-1}. 
%Namely, one has
% \begin{align}
%	& 	x =\dfrac{r}{a} \label{stationarycondition1_d1_4}\\
%	& y = \dfrac{1+r^{\yesp\rodr}}{1+r^{\rodr\nop\yesp}} 
%	\label{stationarycondition1_d1_5} \\
%	&  a^{\yesp\rod} = r^{\yesp\rodr}\dfrac{x}{1+x}-\dfrac{1}{1+x} \\
%	& a^{\rod\yesp} = r^{\yesp\rodr}\dfrac{1}{1+x} -\dfrac{x}{1+x}\\
%	& a^{\yesp\rod\yesp} = -r^{\yesp\rodr} \\
%	& a^{\yesp\nop\rod} = r^{\rodr\nop\yesp}\dfrac{1}{1+x}  \\
%	& a^{\rod\nop\yesp} = r^{\rodr\nop\yesp}\dfrac{x}{1+x}.
%	\label{end_d1_6}
%\end{align}
%Thus, the stationary distribution of the process in this case becomes
%\begin{equation}
%		\bar{\pi}(\boldsymbol{\eta}) = 
%		\dfrac{1}{\bar{Z}}\left(\dfrac{r}{a}\right)^{\sum_{i=1}^{N}-3/2+ 
%		\alpha_i}\left( 
%		\dfrac{1+r^{\yesp\rodr}}{1+r^{\rodr\nop\yesp}} 
%		\right)^{-\sum_{i=1}^{N}\delta^L_{k_{i+1}, k_i+\ell}}
%	\end{equation}
%	where $\bar{Z}$ is the partition function.
%
%
\section{Results and Discussion}
%\section{Properties of the RNAP model}

Given an average density $\rho=N/L$ of rods of the lattice, a central quantity
of interest are the statistical properties of the distance between rods, 
expressed in terms of the headway $m_i$ which is the number of 
empty sites 
between neighboring rods $i^{th}$ and $(i+1)^{th}$.  This quantity,
apart from its intrinsic interest, also determines further important properties
of the stationary translocation kinetics, in particular the stationary flux
related to the rate of elongation and the average excess of bound and 
unbound RNAP.
These quantities are computed below.
 It is not surprising that some 
formulas in the present work turn out to be resemble corresponding 
expressions in
\cite{Belitsky2019-1,Belitsky2019-2} since the Boltzmann factor 
\eqref{invarmeas_d1} is of similar form as in those papers. However, the
parameter $x$ in our setting depends not only on the rates $r, a$ (as in 
\cite{Belitsky2019-1}) but also on the rates $\ell, d$. Moreover, the value 
$y$ in this work is also different from the one in \cite{Belitsky2019-1} since it 
depends on the parameters $r^{\yesp\rodr}, r^{\rodr\nop\yesp}$ and $ 
\ell^{\rodl\yesp}, \ell^{\yesp\nop\rodl}$ characterizing translocation and 
backtracking, respectively, while the same value in 
\cite{Belitsky2019-1,Belitsky2019-2} depends only the forward translocation.

\subsection{Average excess}
The simplest measure that characterizes the distribution of RNAP is the 
average excess density with no PP$_i$ bound over the PP$_i$ bound state 
of RNAP given by
\begin{equation}
	\sigma = \dfrac{\left< N^1 \right> - \left< N^2 \right>}{L}.
\end{equation}
where $N^\alpha$ is the number of rods in state $\alpha$.
For a configuration $\boldsymbol{\eta}$ with $N$ rods
one has by definition $N^1(\boldsymbol{\eta}) + N^2(\boldsymbol{\eta}) = N$.
With the second equality in the definition \eqref{Bdef} the factorization of the 
Boltzmann weight \eqref{boltzmann_d1} in the
invariant measure (\ref{invarmeas_d1}) into an interaction part and the excess
part with the Lagrange multiplier $\lambda$ thus yields
%\begin{claim}\label{aver_ek_d1}
%	The average excess is computed as follows
	\begin{equation}\label{averexcess_d1}
		\sigma = \dfrac{1-x}{1+x}\rho .
	\end{equation}
%\end{claim}
%\textbf{Average densities of each RNAP state:} 

We denote by
\begin{equation}
	\rho_{\alpha} := \left< \delta_{\alpha_i,\alpha} \right> = \dfrac{1}{L}\left< 
	N^\alpha \right>,\ \ \alpha \in \{1,2\},
\end{equation}
the average densities of rods in states $1, 2$. Since $\rho_{1} 
+\rho_{2} = \rho$, one gets from 
(\ref{averexcess_d1})
%\begin{claim} Average densities of each RNAP state are as follows
	\begin{equation}\label{statedensity_d1}
	\rho_{1} =\dfrac{1}{1+x}\rho, \ \ \rho_{2} = \dfrac{x}{1+x}\rho.
\end{equation}
The prefactors
\begin{equation}
\tau_1 := \dfrac{1}{1+x}, \quad \tau_2 := \dfrac{x}{1+x}
\label{tau12def}
\end{equation}
appearing in the consistency relations for the interaction parameters
\eqref{end_d1_1}
thus play the role of the fraction of RNAP in states 1 and 2 respectively.
Correspondingly,
\begin{equation}
x = \frac{\rho_2}{\rho_1}
\end{equation}
is the stationary ratio of RNAP in states 1 and 2.  According to the consistency 
relation \eqref{stationarycondition1_d1_1}. This quantity depends
only on the bare rates $r,\ell,a,d$ not on the interaction parameters.
With the empirical values 
of $r, \ell, a$, $d$ as in \eqref{parameters_d1} one finds $x=31.95$.

\subsection{Absence of headway correlations}

As a first result we note that the headways between rods are uncorrelated.
To prove this we note that
due to translation 
invariance, an allowed configuration of RNAPs can be specified by the 
headway vector $\textbf{m}:= (m_1,...,m_N)$ and the state vector 
$\boldsymbol{\alpha} = 
(\alpha_1,...,\alpha_N)$. Thus, one has $m_i = 
k_{i+1} - (k_i + l_{rod}) \mod L$ and the total number of vacant sites is $M= L 
- 
l_{rod} 
N$. We denote by 
\begin{equation}\label{newvar_d1}
	\theta_{i}^p := \delta_{m_i, p} = \delta_{k_{i+1}, k_i + l_{rod} +p}
\end{equation}
the indicator functions on a headway of length $p$ (in units of base pair) with 
the index $i$ taken modulo $N$, i.e., $\theta_0^p \equiv \theta_N^p$. In 
terms of the parameters \eqref{xydef} and the new distance 
variables \eqref{newvar_d1} one rewrites the stationary distribution 
\eqref{invarmeas_d1} as follows
\begin{equation}\label{invmeas2_d1}
	\tilde{\pi}(\boldsymbol{\zeta}) =\dfrac{1}{Z}\prod_{i=i}^{N}\left( 
	x^{-3/2+\alpha_i}y^{-\theta_i^0}\right)
\end{equation}
where $\boldsymbol{\zeta}$ is an allowed configuration defined by state 
vector $\boldsymbol{\alpha}$ and headway vector $\textbf{m}$. Notice that 
the measure \eqref{invmeas2_d1} is of factorized form which indicates the 
absence of headway correlations.

As in \cite{Belitsky2019-1,Belitsky2019-2}, we work in the grand-canonical 
ensemble defined by
\begin{equation}\label{gc_d1}
	\tilde{\pi}_{gc}(\boldsymbol{\zeta}) = 
	\dfrac{1}{Z_{gc}}\prod_{i=1}^{N}(x^{-3/2+\alpha_i} 
	y^{-\theta_{i}^0}z^{m_i}),
\end{equation}
where $Z_{gc} = (Z_1Z_2)^N$ with
\begin{equation}
	Z_1 = \dfrac{1+(y-1)z}{y(1-z)},\ \ Z_2 = x^{1/2} + x^{-1/2},
\end{equation}
and the solution of the quadratic equation
\begin{equation}
(y-1) z^2 + z \left(
y \frac{1-(l_{rod}-1)\rho}{1-l_{rod}\rho}-2(y - 1) \right) - 1 = 0
\end{equation}
given by
\begin{equation}\label{value_z_d1}
	z:=z(\rho,y)= 1 -\dfrac{1 - (l_{rod}-1)\rho - \sqrt{(1 - (l_{rod}-1)\rho)^2 - 
	4\rho(1-l_{rod}\rho)(1-y^{-1}) }}{2(1-l_{rod}\rho)(1-y^{-1})}.
\end{equation}
which parametrizes the density of rods.
%As shown in \cite{Belitsky2019-1}, the measure \eqref{gc_d1} is well-defined.
In the absence of nearest-neighbor static interaction, i.e., when RNAP only experience excluded volume interaction, this relation reduces
to 
\begin{equation}
z_0 := z(\rho,1) = \frac{1-l_{rod}\rho}{1- (l_{rod}-1)\rho}.
\end{equation}
By definition, for any static interaction strength the mathematically maximal density 
of rods is $\rho_{max} = l_{rod}^{-1}$ which expresses full coverage of the 
lattice 
by rods. For all $y$ one has $z(0,y) = 1 \geq z(\rho,y) \geq 0 = 
z(\rho_{max},y)$. Hence for rod densities of interest, i.e., $\rho \neq 0, 
\rho_{max}$ one has $0<z<1$ and $z$ is strictly monotonically decreasing in 
$\rho$.

\subsection{Headway distribution}

Since the invariant measure is of the same form as in \cite{Belitsky2019-1, 
Belitsky2019-2}, mean headway and headway distribution are the same form 
as in those papers as functions of the parameters $x,y,z$, the difference 
being the dependence of these parameters on the microscopic transition
rates (\ref{stationarycondition1_d1_1}), (\ref{stationarycondition1_d1_2}), and the density parameter (\ref{value_z_d1}).

Denote by $P_h(r)$ the distribution of the headway between the front of a 
trailing rod $i$ and the back of a leading rod $i+1$ which means $P_h(r) = 
\dfrac{1}{\rho}\left< \delta_{k_{i+1} - k_i - l_{rod}, r}\right> = \left< \theta_i^r 
\right>, r \in \mathbb{N}$ where $\mathbb{N} = \{0,1,2...\}$ are the natural
numbers. This distribution depends on the rod density $\rho$ and the 
interaction
parameter $y$. However, to keep notation light we omit this dependence.
From \eqref{gc_d1} one finds
	\begin{equation}\label{meanheadway_d1_1}
P_h(r) = 
		\begin{cases}
			\dfrac{1-z}{1+(y-1)z} \ \ \text{ for } r = 0, \\
			yP_h(0)z^r \ \ \text{ for } r \geq 1
		\end{cases}
	\end{equation}
with the mean headway
	\begin{equation}\label{meanheadway1_d1_1}
	\bar{\lambda} (\rho) :=	\left<m_i \right> = \dfrac{yz}{(1-z)(1+(y-1)z)} = 
	\dfrac{1-l_{rod}\rho}{\rho}
	\end{equation}
%\end{claim}
in units of the lattice constant $\delta$ given by the size of a DNA base pair.
The mean headway does not depend on the static interaction 
strength.

The nearest-neighbor probability
\begin{equation}
p_0 := P_h(0) = \dfrac{1-z}{1+(y-1)z}
\label{p0}
\end{equation}
of having headway 0 plays a special role. This is the probability of finding
two rods as nearest neighbors which determines
the mean static interaction energy density
	\begin{equation}
\label{meanenergy1_d1_1}
	\bar{u} (\rho,y) := \frac{1}{L J} \sum_{\mathbf{k}} \left<U(\mathbf{k})\right> 
=  \frac{1}{N} \sum_{i=1}^{N} \left< \theta_i^0 \right> = p_0
%= 	\dfrac{1-z}{1+(y-1)z} 
	\end{equation}
in units of the interaction constant $J$.
%
% We rewrite them here through claims \ref\boldsymbol{\alpha} and 
% \ref{head_dist_d1} without proving them.
%\begin{claim}\label\boldsymbol{\alpha}

\begin{figure}[H]
	\centering
	\begin{subfigure}[b]{0.43\textwidth}
		\centering
		\includegraphics[width=\textwidth]{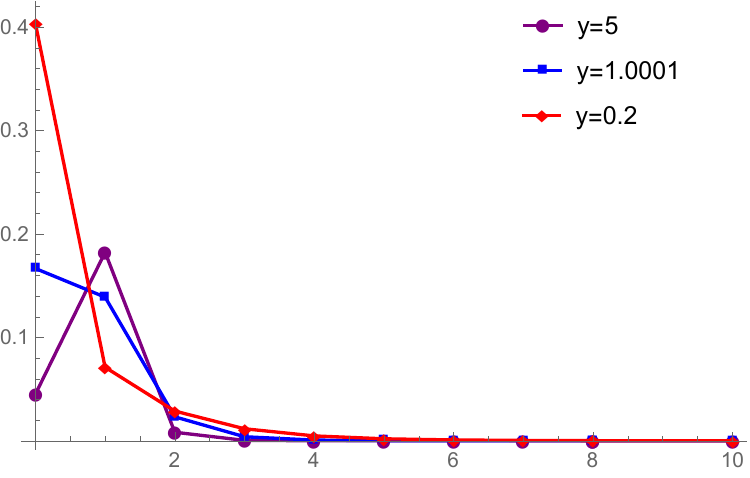}
		\caption{RNAP density $\rho = 0.1$.}
		\label{fig_headway_d1_0.1}
	\end{subfigure}
	\begin{subfigure}[b]{0.43\textwidth}
		\centering
		\includegraphics[width=\textwidth]{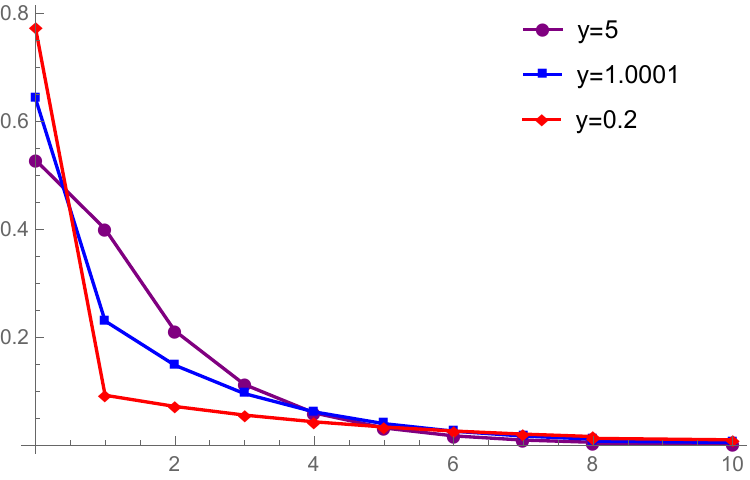}
		\caption{RNAP density $\rho = 0.18$.}
		\label{fig_headway_d1_0.18}
	\end{subfigure}
	\caption{RNAP headway distribution $P_h(r)$ for different static interaction 
	strengths $y$ as a function of the integer lattice distance at average RNAP 
	densities $\rho = 0.1$ and $\rho= 0.18$. The connecting lines are guides to 
	the eye.}
\label{fig_headway_d1}
\end{figure}

%\begin{equation}
%z_0 = \frac{\bar{\lambda}}{1 + \bar{\lambda}}.
%\end{equation}
%
%\begin{equation}\label{value_z_d2}
%	z = 1 -\dfrac{1+\bar{\lambda} - \sqrt{(1+\bar{\lambda})^2 - 
%	4 \bar{\lambda}(1-y^{-1}) }}{2 \bar{\lambda}(1-y^{-1})} = 
%1 -\dfrac{1- \sqrt{1 - 
%	4 z_0(1-z_0) (1-y^{-1}) }}{2 z_0 (1-y^{-1})} 
%\end{equation}
%\begin{equation}
%P_h(0) = \dfrac{(1-z)^2}{(1-z)(1+(y-1)z)} = \dfrac{(1-z)^2}{yz} \bar{\lambda}
%\end{equation}
%
%
To examine the impact of the static interaction strength parametrized by $y$ on the 
full distribution we first note that $P_h(r)$ is strictly decreasing for headways 
$r 
\geq 1$ independently of interaction 
strength and rod density. This is an entropic effect which indicates that the 
number of allowed configurations decreases with the headway
between them. However, the probability $p_0$ of 
finding two neighboring rods depends non-trivially on the interplay of 
interaction strength and rod density. 
Since $p_0 = P_h(1)/(yz)$ and since $z<1$, any attractive interaction
(which corresponds to $y<1$) 
yields $p_0 > P_h(1)$ which is indeed expected for attraction. However,
somewhat contrary to intuition, 
the probability of finding two rods as immediate neighbors is smaller than the 
probabilty of finding them with an empty site between them even for repulsive
interaction as long as it is not too strong.
Only above a critical repulsive interaction strength that depends on the density 
through the relation $y>1/z$ the next-nearest neighbor headway probability 
$P_h(1)$ exceeds the nearest neighbor headway probability $p_0$. 
This effect is demonstrated in Fig. \ref{fig_headway_d1}
 for rods of length $l_{rod}=5$ for two different rod densities and three 
 different 
 interaction
parameters $y=0.2$ (attractive static interaction), $y=1.0001$ (very weak
repulsive interaction), and $y=5$ (strong repulsive static interaction).
 Since backtracking is reduced when there are many neighboring rods we 
 conclude that backtracking due to neighbor depletion sets in for strong 
 repulsive static interactions above the critical value $y_c=1/z$.

%\begin{remark}
%	Since $z<1$ then function $P_h(r)$ decreases, when $r \geq 1$. 
%Consequently, in order to compare 
%	$P_h(0)$ with $P_h(1)$ 	(expressed in the distribution 
%\eqref{meanheadway_d1_1}), one only needs to
%	compare $yz$ with $1$: if $yz>1$, then $P_h(1) > P_h(0)$. One can see 
%from
%	Fig.~\ref{fig:yz_attaction_d1} that $yz <1$, thus for the attractive case 
%$0<y<1$ one has 
%	$P_h(1) < P_h(0)$. Also, examining the Fig. \ref{fig:yz_repulsion_d1} we 
%see that at high density, 
%	even for very strong repulsion ($y$ large), it holds that $yz <1$ which 
%implies in this case that
%	$P_h(1) < P_h(0)$ (Figs. \ref{fig_headway_d1_0.18}). However,  
%	Fig.~\ref{fig:yz_repulsion_d1} 	also shows that at low density occurs the 
%following phenomenon: 
%	if $y$ is close to 1, one has $yz < 1$ while if $y$ is large enough, one has 
%$yz >1$, and hence 
%	one can compare $P_h(1)$ with  $P_h(0)$ 
%(Fig.~\ref{fig_headway_d1_0.1}).
%\end{remark}

\subsection{Average elongation rate}\label{elong_rate_d1}

The main quantity of interest is the average elongation rate which is
related to the flux of rods along the chain. To elucidate the effect of 
backtracking we first compute for the mean velocity of a 
single RNAP which experiences no interaction with another RNAP
and how it changes qualitatively if we assume a rate of backtracking $\ell$
different from the empirical value reported in \cite{Wang1998}.

\subsubsection{Mean velocity of a single RNAP} 

For a single rod the process reduces to a biased random walk of a particle
with an internal degree of freedom that is given by the two chemical states
ini which the RNAP can be.
Following the approach of Wang et al \cite{Wang1998}, one finds by
straightforward computation
\begin{equation}
\label{velocityoneRNAP_d1}
	v_0  = \dfrac{ra - \ell d}{r + \ell + a + d} 
= r \tau_1 - \ell \tau_2 
= \frac{r - \ell x}{1+x}
\end{equation}
If an RNAP would perform a simple random walk, then its velocity would be
$v_0 = r - \ell$ which differs from
	\eqref{velocityoneRNAP_d1} by the prefactors $\tau_1=\dfrac{\rho_1}{\rho}, 
	\tau_2=\dfrac{\rho_2}{\rho}$ which are the number fraction of the 
	chemical states 
	1 and 2, respectively. This difference quantifies the effect of the mechanochemical
cycle on the average velocity $v_0$ of an RNAP.
At low density $\rho$ of RNAP, i.e., in a scenario when RNAP would almost 
never become neighbours on the DNA strand, the RNAP flux is then given by
\begin{equation}
j_0 = \rho v_0.
\label{noninteractingcurrent}
\end{equation}

We can also read off the effect of backtracking on the velocity of a single 
RNAP. To quantify this effect we denote by $v_0^{ref}$ the hypothetical
velocity in the
absence of backtracking ($\ell=0$) and by $v_0(10)$ the velocity for
strong backtracking  for which we take a tenfold backtracking rate
$10 \ell$ compared to the empirical rates reported in \cite{Wang1998}. This yields
%with the choice \eqref{parameters_d1} 
the ratios
%$r=10^3$ $\ell=0.21$ $a=31.4$ $d=10$
%\begin{equation}
%v_0 = \dfrac{31397.9}{1041.61} = 30.14
%v_0^{ref} = \dfrac{31400}{1041.4} = 30.15 
%v_0(10) = \dfrac{31190}{1062.4} = 29.36
%\end{equation}
\begin{equation}
\frac{v_0}{v_0^{ref}} = 0.9997, \quad
\frac{v_0(10)}{v_0^{ref}} = 0.974.
\end{equation}
Hence assuming a complete absence of backtracking yields no perceptible
change in the avarage velocity of a single RNAP. The reduction of the
velocity for strong backtracking compared to $v_0^{ref}$ is small (about 
$2.6\%$) even though
the backtracking rate has been taken as 10 times the empirical rate $\ell$.
This observation leads us to conclude that the effect of backtracking on the 
velocity of single RNAP at, at most, small.

%=================================

\subsubsection{Average velocity}
\label{a_velocity_flux} 

To elucidate how the interplay of backtracking with the static and kinetic
interactions affects the average elongation rate we investigate the average 
velocity $v$ of an RNAP in an ensemble of interacting RNAP, as opposed to 
the single RNAP discussed above. The stationary average flux 
\begin{equation}
	j =\rho v.
\end{equation}
of RNAP along the DNA template is the average number of RNAP crossing 
a lattice bond per unit time (second) \cite{Tripathi2008} is thus a measure of 
the average elongation rate.

In the framework of our model, $j$ is given by the expectation of the right 
and left rod jump rates $r_i(\boldsymbol{\eta})$, $\ell_i(\boldsymbol{\eta})$ 
with respect to the stationary distribution (\ref{invarmeas_d1}) through the 
difference $j= \langle r_i - \ell_{i}\rangle$. From the 
definitions \eqref{rate1_d1}, \eqref{rate2_d1} one has
\begin{equation}
	j  = \frac{r}{L} \langle N_1 \left( 1 + r^{\yesp\rodr} \theta_{i-1}^0\right) 
	\left(1-\theta_i^0  \right)  + 	r^{\rodr\nop\yesp} \theta_i^1\rangle - 
 \frac{\ell}{L}  \langle N_2 \left(1 + \ell^{\rodl\yesp} \theta_i^0 \right) 
 \left(1-\theta_{i-1}^0\right)  + \ell^{\yesp\nop\rodl}	\theta_{i-1}^1\rangle 
\label{jstat}
\end{equation}
where we have used that $\delta_{k_{i}+l_{rod}, k_{i+1}}
\cdot \delta_{k_{i}+l_{rod}+1,k_{i+1}} = \delta_{k_{i-1}+l_{rod}, k_{i}}
\cdot \delta_{k_{i-1}+l_{rod}+1,k_{i}} =0$.
The expectation \eqref{jstat} does not depend on the rod $i$ because 
of stationarity and the conservation of the total number of rods during translocation. 

The factorization property of the stationary distribution 
\eqref{invmeas2_d1} allows for expressing the expectation of the 
products appearing in this formula by the product of expectations
involving the stationary headway probabilities 
$\langle \theta_{i}^0 \rangle = P_h(0) = p_0$, $\langle \theta_{i}^1 \rangle 
= P_h(1) = yz p_0$ given by \eqref{meanheadway_d1_1}.
The headway distribution \eqref{meanheadway_d1_1} 
and the consistency relation \eqref{stationarycondition1_d1_2} then yields
\begin{align}
%j	& = r \rho_{1} \left[ \left( 1+r^{\yesp\rodr} p_0 \right)  \left(1-p_0\right) 
%       + r^{\rodr\nop\yesp} yz p_0 \right]
%    - \ell \rho_{2}  \left[ \left(1 + \ell^{\rodl\yesp}  p_0\right) 
%    \left(1-p_0\right) 
%    + \ell^{\yesp\nop\rodl} yz p_0 \right]
%\label{fluk_d1} \\
j	& = j_0 z \left[1 + \left( r \rho_{1} r^{\yesp\rodr} 
- \ell \rho_{2} \ell^{\rodl\yesp} \right) 
(1-z)\dfrac{y + 1+(y-1)z}{[1+(y-1)z]^2}  \right]
\label{j2_d1} \\
%v	& = r \tau_{1} \left[ \left( 1+r^{\yesp\rodr} p_0 \right)  \left(1-p_0\right) 
%       + r^{\rodr\nop\yesp} yz p_0 \right]
%    - \ell \tau_{2}  \left[ \left(1 + \ell^{\rodl\yesp}  p_0\right) 
%    \left(1-p_0\right) 
%    + \ell^{\yesp\nop\rodl} yz p_0 \right]
%\label{v_d1} \\
v	
& = v_0 z \left[1+ \left( r \tau_{1} r^{\yesp\rodr} 
- \ell \tau_{2} \ell^{\rodl\yesp} \right) 
(1-z)\dfrac{y + 1+(y-1)z}{[1+(y-1)z]^2} \right] %\\
%& = v_0 z + \left( r \tau_{1} r^{\yesp\rodr} 
%- \ell \tau_{2} \ell^{\rodl\yesp} \right) y^{-1}
%(1-z)\dfrac{1 + (y-1) z \left(1 - 
%\frac{\rho}{1-l_{rod}\rho}  \right)  + z \left(y - 
%\frac{\rho}{1-l_{rod}\rho}  \right) }{1 + (y-1)z \left(1 -
%\frac{\rho}{1-l_{rod}\rho} \right) }  \\
\label{v2_d1} 
\end{align}
in terms the kinetic interaction parameters for pushing.

To discuss the effect of the microscopic interactions on the collective 
behaviour of a stationary ensemble
of RNAP in terms of the average RNAP velocity \eqref{v2_d1}
we define the {\it interaction factor}
\begin{equation}
q := \frac{v}{v_0} = \frac{j}{j_0} =
z \left[ 1 +  \frac{ r \tau_{1} r^{\yesp\rodr} 
- \ell \tau_{2} \ell^{\rodl\yesp}}{r \tau_{1} 
- \ell \tau_{2} }
(1-z)\dfrac{y + 1+(y-1)z}{[1+(y-1)z]^2}  \right]
\label{qdef}
\end{equation}
which quantifies how much the velocity (or flux) is affected by the presence
of RNAP interactions.
Notice that $q$ depends both on the average RNAP density $\rho$
an the various rates and interaction parameters that define the microscopic
interactions between individual RNAP. Thus $q$ characterizes whether 
the average velocity of an interacting system of
RNAP is enhanced or reduced compared to a hypothetical system of 
noninteracting RNAP that is effectively described by translocation at very low 
RNAP density. We speak of boosting when $q>1$ for a range of RNAP
densities and system parameters and of jamming when $q<1$. 

%\begin{eqnarray}
%v 
%& = &  r \tau_{1}  \left( 1+r^{\yesp\rodr} P_h(0) \right)  \left(1-P_h(0)\right) 
%    - \ell \tau_{2}  \left(1 + \ell^{\rodl\yesp}P_h(0)\right) \left(1-P_h(0)\right) 
%\nonumber \\
%& = &  r \tau_{1}  \left( 1+ (y-1) P_h(0) \right)  \left(1-P_h(0)\right) 
%    - \ell \tau_{2}  \left(1 + (y-1) P_h(0)}\right) \left(1-P_h(0)\right) 
%\nonumber \\
%& = &  v_0  \left( 1+ (y-1) P_h(0) \right)  \left(1-P_h(0)\right) 
%\nonumber \\
%& = &  v_0  \left( 1+ (y-1)\frac{1-z}{1+(y-1)z} \right)  
%\left(1-\frac{1-z}{1+(y-1)z}\right) 
%\nonumber \\
%& = &  v_0  \left(  \frac{1+(y-1)z
%+ (y-1)(1-z)}{1+(y-1)z} \right)  
%\left(\frac{1+(y-1)z -1+z}{1+(y-1)z}\right) 
%\nonumber \\
%& = &  v_0  \left(  \frac{y}{1+(y-1)z} \right)  
%\left(\frac{yz}{1+(y-1)z}\right) 
%\nonumber \\
%\end{eqnarray}
%
%
%
%%One notices that Figures \ref{fig:vj_5_0_d1} and \ref{fig:vj_5_9_d1} are 
%%similar since the rate of jump backward is small comparing to the jump 
%%forward 
%%rate. So that, in the following figures, we only plot average velocity and flux 
%%with $\ell^{\yesp\nop\rodl} = -0.9$.
%
%

\subsection{Role of Backtracking for boosting}\label{discussion}

To examine the relationship between microscopic backtracking and the
collective behaviour that leads to boosting 
we work with the length of RNAP $l_{rod} =5$ and use the values 
of $r, \ell, a$, $d$ given in \eqref{parameters_d1} and below.
Kinetic interaction strengths are varied in different ways and the effect 
on boosting is discussed
for the full range of rod densities, ranging from 0 to 
the maximal rod density $1/\ell=0.2$

\subsubsection{Minimal kinetic interaction range} 

It is interesting that if 
$r^{\rodr\nop\yesp} = \ell^{\yesp\nop\rodl} = 0$ (minimal kinetic interaction 
range), from identity \eqref{stationarycondition1_d1_2} one has 
$r^{\yesp\rodr} = \ell^{\rodl\yesp}=y-1$. Then the average velocity takes the 
simple 
form
%\begin{eqnarray}
%v 
%& = &  r \tau_{1}  \left( 1+r^{\yesp\rodr} P_h(0) \right)  \left(1-P_h(0)\right) 
%    - \ell \tau_{2}  \left(1 + \ell^{\rodl\yesp}P_h(0)\right) \left(1-P_h(0)\right) 
%\nonumber \\
%& = &  r \tau_{1}  \left( 1+ (y-1) P_h(0) \right)  \left(1-P_h(0)\right) 
%    - \ell \tau_{2}  \left(1 + (y-1) P_h(0)}\right) \left(1-P_h(0)\right) 
%\nonumber \\
%& = &  v_0  \left( 1+ (y-1) P_h(0) \right)  \left(1-P_h(0)\right) 
%\nonumber \\
%& = &  v_0  \left( 1+ (y-1)\frac{1-z}{1+(y-1)z} \right)  
%\left(1-\frac{1-z}{1+(y-1)z}\right) 
%\nonumber \\
%& = &  v_0  \left(  \frac{1+(y-1)z
%+ (y-1)(1-z)}{1+(y-1)z} \right)  
%\left(\frac{1+(y-1)z -1+z}{1+(y-1)z}\right) 
%\nonumber \\
%& = &  v_0  \left(  \frac{y}{1+(y-1)z} \right)  
%\left(\frac{yz}{1+(y-1)z}\right) 
%\nonumber \\
%\end{eqnarray}
%
\begin{equation}\label{velocity_d11}
v =  v_0 z \left(\dfrac{y}{1+(y-1)z}\right)^2
\end{equation}
where $z$ given by \eqref{value_z_d1} is a function of the RNAP density
and static interaction strength. Thus the interaction factor \eqref{qdef}
\begin{equation}
q = z \left(\dfrac{y}{1+(y-1)z}\right)^2 %= (\bar{\lambda})^2 z^{-1} (1-z)^2
\label{q_d11}
\end{equation}
is a function only of the RNAP density and the interaction terms of the model
and can be expressed in terms of the density and the static interaction
alone. 

\begin{figure}[H]
	\centering
	\begin{subfigure}[b]{0.4\textwidth}
		\centering
		\includegraphics[width=\textwidth]{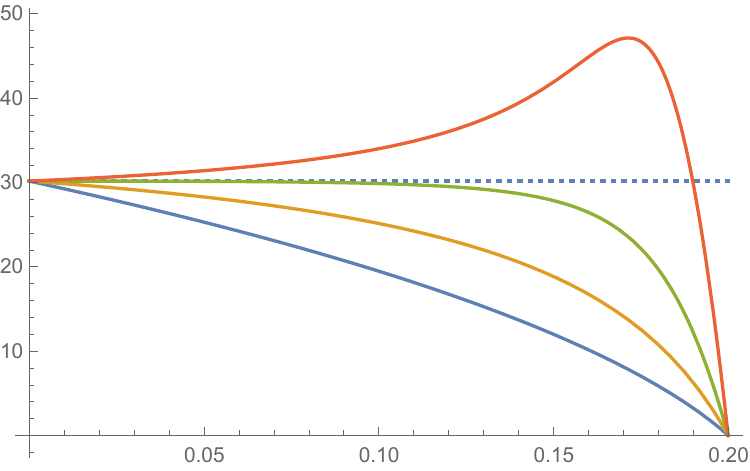}
		\caption{Average velocity $v$}
		\label{fig:v_0_0_d1}
	\end{subfigure}
	\begin{subfigure}[b]{0.4\textwidth}
		\centering
		\includegraphics[width=\textwidth]{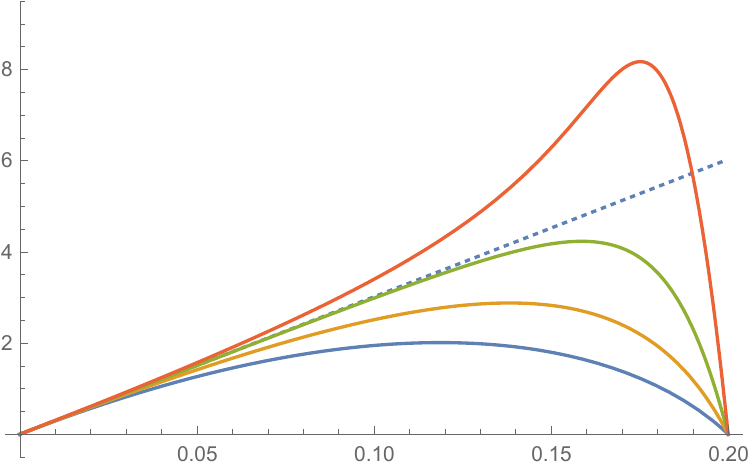}
		\caption{Average flux $j$}
		\label{fig:j_0_0_d1}
	\end{subfigure}
	\caption{RNAP velocity $v$ and RNAP flux $j$ as functions of RNAP 
	density with $r^{\rodr\nop\yesp} = \ell^{\yesp\nop\rodl} =0$ and  
	different static interaction strengths $y$. Curves from top to bottom with value of 
	$y: 5, 2, 1.001, 0.5$. The dotted lines correspond to 
	non-interacting RNAP.}
	\label{fig:vj_0_0_d1}
\end{figure}

The result \eqref{velocity_d11} for the average velocity shows that
backtracking manifests itself through the bare backtracking rate $\ell$ in the 
overall amplitude $v_0$ given by the velocity of an isolated RNAP. 
However, as discussed 
above, the effect is so small that curves for different values of
$\ell$ would collapse onto the same curves shown in Fig. \ref{fig:vj_0_0_d1}
for $\ell$ given by \eqref{parameters_d1}. Moreover, this particular
effect of backtracking has no impact on boosting.

To examine backtracking affects on boosting through the RNAP interactions 
we first note that by definition, $q=1$ at density $\rho=0$.
As a function of the density, the interaction factor has a maximum at a density $\rho^\ast$ given by $z =1/(y-1)$ where the derivative of w.r.t. the density vanishes.
Since $0\leq z < 1$, this can happen only if $y>2$
which implies that the velocity increases and reaches a maximum 
$v^\ast>v_0$ only for a sufficiently strong static repulsion
given by the critical value
\begin{equation}
y_c=2.
\label{yc}
\end{equation}
Above $\rho^\ast$ the interaction factor decreases and boosting disappears
at a critical density $\rho_c$ given by the nonzero solution of the equation $q(\rho_c,y)=1$.
Thus for $y > y_c$ boosting occurs in the density range $0 < \rho < \rho_c$
as shown in Fig. \ref{fig:vj_0_0_d1}. In the high density range $\rho_c < \rho 
\leq 1/l_{rod}$ jamming takes over.

The same observation was made in \cite{Belitsky2019-1} in the absence of
backtracking and shows that the emergence of
boosting arises by the same interplay between static interaction and
kinetic interactions even when the interaction parameter $\ell^{\rodl\yesp}$
for backtracking is as strong as the interaction parameter 
$r^{\yesp \rodr}$ for translocation. Thus, in the minimal kinetic interaction 
range, only for
sufficiently strong static repulsion $y>2$ boosting appears and 
reaches global maximum at a density
$\rho^\ast$, above which the velocity drops from the maximum to zero
at the maximal rod density $0.2$. Above the critical static repulsion strength
boosting thus occurs for densities between 0 and a critical density
that is close to the maximally possible density $1/l_{rod}$.
When $y$ is not strong enough ($y \leq 2$), the velocity is less than the 
velocity of a 
single RNAP (dotted line). See Fig. \ref{fig:v_0_0_d1} for these features and 
Fig. \ref{fig:j_0_0_d1} that shows how the corresponding the average flux $j$ 
varies with the density.  The interaction factor $q$ is given by the same curves 
as in Fig. \ref{fig:v_0_0_d1}, with rescaled $y$-axis where the dotted reference
line for non-interaction RNAP at $y=1$.

The interaction parameter $\ell^{\rodl\yesp}=r^{\yesp\rodr}$, related to both 
backtracking and translocation, however, has a 
significant impact on boosting
as it is linked with the static interaction strength via
the consistency relation $\ell^{\rodl\yesp}=y-1$ which shows that 
boosting arises if and only if $\ell^{\rodl\yesp}=r^{\yesp\rodr}>1$. This means 
that for boosting to emerge 
it is not enough that the RNAP pushing on the level of individual RNAP, 
which corresponds to $\ell^{\rodl\yesp}_c=r^{\yesp\rodr}_c >0$, exists. It 
has to be sufficiently strong and exceed the critical value determined by the interaction parameter
$\ell^{\rodl\yesp}_c=r^{\yesp\rodr}_c=1$. Moreover, when sufficiently strong it 
is the RNAP pushing
itself that is important, not its direction. 
Hence backtracking has no significant impact on the rate of elongation in the 
scenario of minimal kinetic interaction range.

To go beyond the basic minimal interaction scheme we account in what 
follows for next-nearest neighbor interaction (extended kinetic interactions).

\subsubsection{Extended kinetic interaction range} 

The picture changes somewhat for extended kinetic interaction range.
Using the exact expressions \eqref{v2_d1} for the average velocity in an
ensemble of interacting RNAP and
\eqref{velocityoneRNAP_d1} for a single rod as well as the consistency 
conditions
\eqref{stationarycondition1_d1_2} one finds that 
the interaction parameter is larger
than 1 in a density range is given by the inequality
\begin{equation}
z  \frac{ 1 + (y-1) z + y}{ \left( 1+(y-1)z\right)^2} < 
\frac{r \tau_{1}  r^{\yesp\rodr} -\ell \tau_{2}   \ell^{\rodl\yesp}}{r \tau_{1}  
 -\ell \tau_{2} }
\end{equation}
relating density (parametrized by $z$) and the static interaction
parameter $y$ to the bare and nearest-neighbor backtracking and 
translocation 
rates. We illustrate this inequality for some scenarios.

\begin{figure}[H]
	\centering
	\begin{subfigure}[b]{0.4\textwidth}
		\centering
		\includegraphics[width=\textwidth]{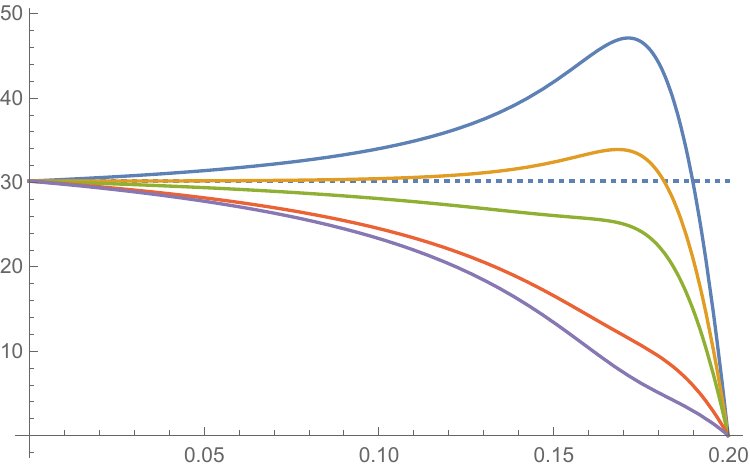}
		\caption{Average velocity $v$}
		\label{fig:v_5_0_d1}
	\end{subfigure}
	\begin{subfigure}[b]{0.43\textwidth}
		\centering
		\includegraphics[width=\textwidth]{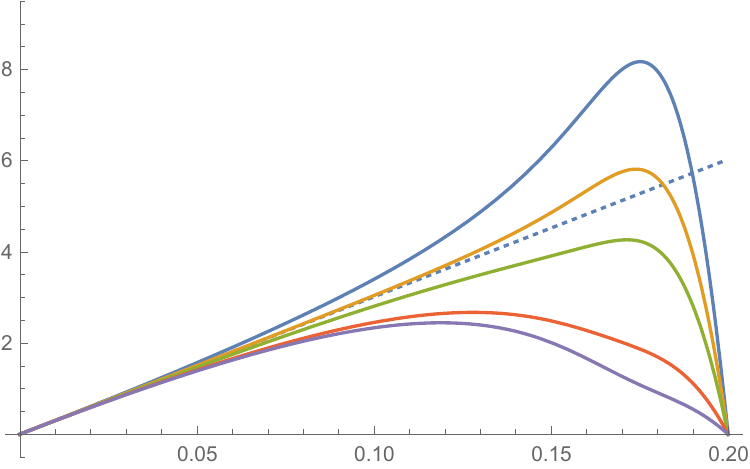}
		\caption{Average flux $j$}
		\label{fig:j_5_0_d1}
	\end{subfigure}
	\caption{RNAP velocity $v$ and RNAP flux $j$ as functions of RNAP density 
	with static interaction strength $y = 5$ and kinetic interaction parameter 
	$\ell^{\yesp\nop\rodl} =0$ for different values $r^{\rodr\nop\yesp}$. Curves 
	from top to bottom with value of $r^{\rodr\nop\yesp}: 0, -0.3, -0.5, -0.8, - 
	0.9$. The dotted lines correspond to non-interacting RNAP.}
	\label{fig:vj_5_0_d1}
\end{figure}

\begin{figure}[H]
	\centering
	\begin{subfigure}[b]{0.43\textwidth}
		\centering
		\includegraphics[width=\textwidth]{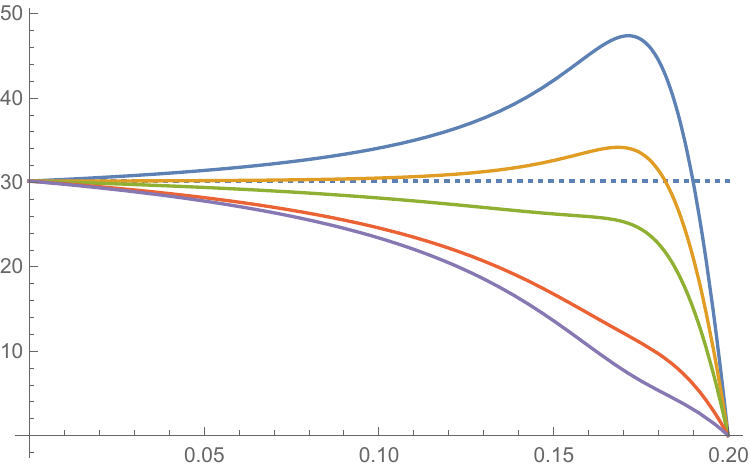}
		\caption{Average velocity $v$}
		\label{fig:v_5_9_d1}
	\end{subfigure}
	\begin{subfigure}[b]{0.43\textwidth}
		\centering
		\includegraphics[width=\textwidth]{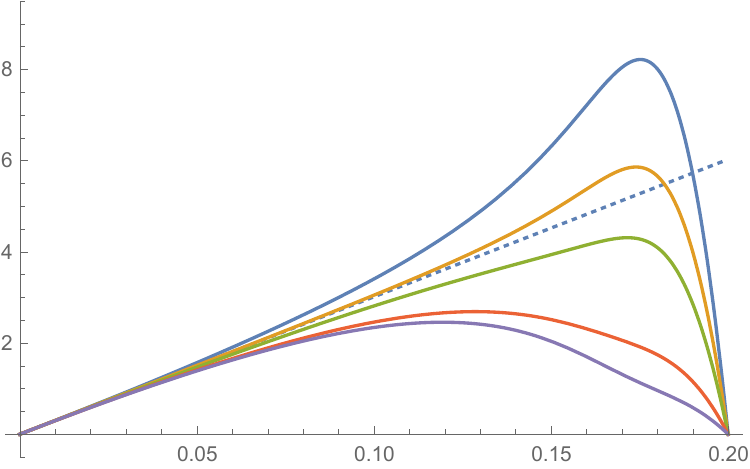}
		\caption{Average flux $j$}
		\label{fig:j_5_9_d1}
	\end{subfigure}
	\caption{RNAP velocity $v$ and RNAP flux $j$ as functions of RNAP density 
	with static interaction strength $y = 5$ and kinetic interaction parameter 
	$\ell^{\yesp\nop\rodl} =-0.9$ for different values $r^{\rodr\nop\yesp}$. 
	Curves from top to bottom with value of $r^{\rodr\nop\yesp}: 0, -0.3, -0.5, 
	-0.8, - 0.9.$ The dotted lines correspond to non-interacting RNAP. }
	\label{fig:vj_5_9_d1}
\end{figure}

\paragraph{Strong static repulsion:} 
First we consider strong repulsion with $y=5$ well
above the critical value for which 
boosting occurs and explore two 
scenarios for backtracking.\\ 
(i) We take $\ell^{\yesp\nop\rodl} =0$ which means
that the next-nearest upstream neighbor has no effect on the rate of RNAP backtracking (no blocking enhancement
for backtracking RNAP). Static repulsion is then
realized by $\ell^{\rodl\yesp} = 4$, i.e., pushing to the
left. In Figs \ref{fig:v_5_0_d1} and  
\ref{fig:j_5_0_d1} it is shown how boosting changes as blocking
enhancement for translocation is increased. When the blocking enhancement for translocation is too 
strong ($r^{\rodr\nop\yesp}$ is close to -1), then even strong 
pushing ($y$ arbitrarily large) does not lead to boosting.\\
(ii) For $\ell^{\yesp\nop\rodl} = -0.9$ the presence of a next-nearest upstream 
neighbor strongly suppresses 
backtracking. For the same choice interaction
parameters for translocation and
the same static repulsion strength.
The curves shown in 
Figs. \ref{fig:v_5_9_d1} and  
\ref{fig:j_5_9_d1} for $\ell^{\yesp\nop\rodl} =-0.9$
are nearly indistinguishable from the curves in Figs. 
\ref{fig:v_5_0_d1} and  
\ref{fig:j_5_0_d1} for $\ell^{\yesp\nop\rodl} =0$.
Hence the occurrence of boosting phenomenon is insensitive to the choice of the phenomenological
static parameter $\ell^{\yesp\nop\rodl}$,
indicating that suppression of boosting by sufficiently strong
blocking enhancement and persistence of of boosting
for low blocking enhancement is robust, also in the
presence of backtracking.

\begin{figure}[H]
	\centering
	\begin{subfigure}[b]{0.43\textwidth}
		\centering
		\includegraphics[width=\textwidth]{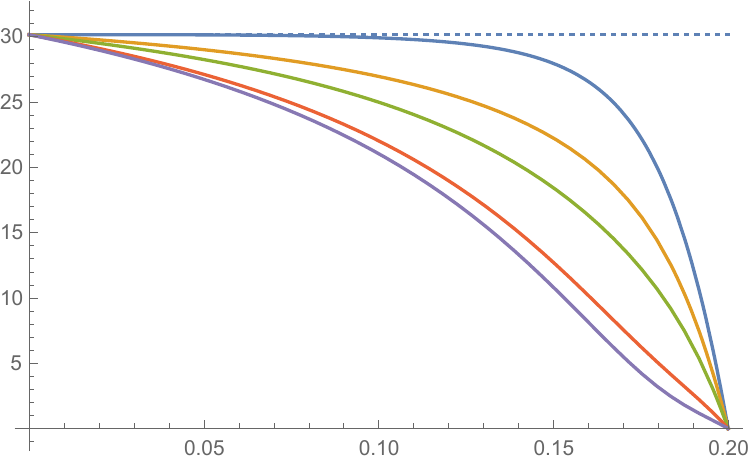}
		\caption{Average velocity $v$}
		\label{fig:v_2_9_d1}
	\end{subfigure}
	\begin{subfigure}[b]{0.43\textwidth}
		\centering
		\includegraphics[width=\textwidth]{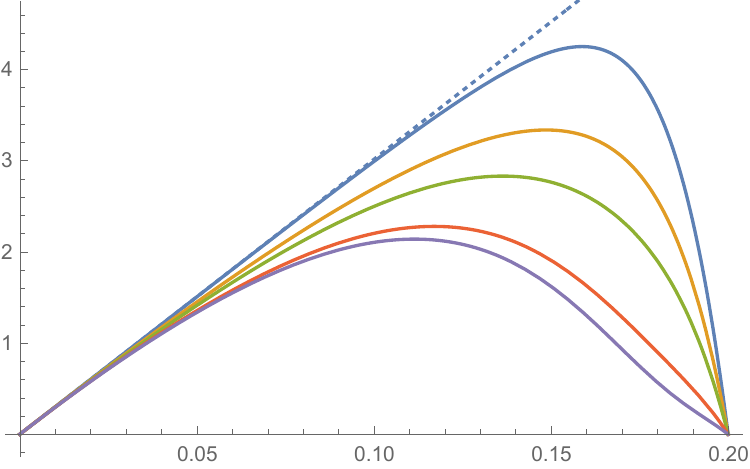}
		\caption{Average flux $j$}
		\label{fig:j_2_9_d1}
	\end{subfigure}
	\caption{RNAP velocity $v$ and RNAP flux $j$ as functions of RNAP density 
	with static interaction strength $y = 2$ and kinetic interaction parameter 
	$\ell^{\yesp\nop\rodl} =-0.9$ for different values $r^{\rodr\nop\yesp}$. 
	Curves from top to bottom with value of $r^{\rodr\nop\yesp}: 0, -0.3, -0.5, 
	-0.8, - 0.9$. The dotted lines correspond to non-interacting RNAP. }
	\label{fig:vj_2_9_d1}
\end{figure}

\begin{figure}[H]
	\centering
	\begin{subfigure}[b]{0.43\textwidth}
		\centering
		\includegraphics[width=\textwidth]{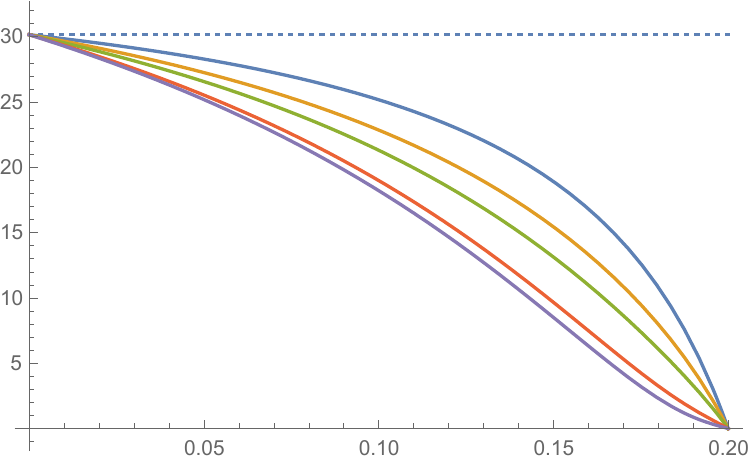}
		\caption{Average velocity $v$}
		\label{fig:v_1_9_d1}
	\end{subfigure}
	\begin{subfigure}[b]{0.43\textwidth}
		\centering
		\includegraphics[width=\textwidth]{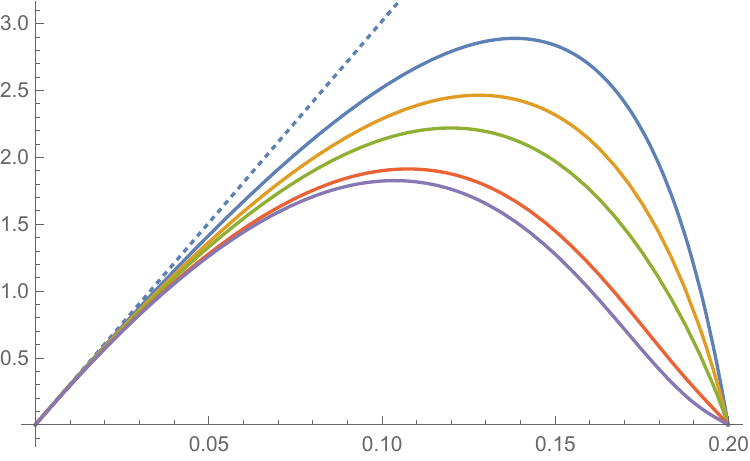}
		\caption{Average flux $j$}
		\label{fig:j_1_9_d1}
	\end{subfigure}
	\caption{RNAP velocity $v$ and RNAP flux $j$ as functions of RNAP density 
	with static interaction strength $y = 1.0001$ and kinetic interaction 
	parameter $\ell^{\yesp\nop\rodl} =-0.9$ for different values 
	$r^{\rodr\nop\yesp}$. Curves from top to bottom with value of 
	$r^{\rodr\nop\yesp}: 0, -0.3, -0.5, -0.8, - 0.9$. The dotted lines correspond 
	to non-interacting RNAP. }
	\label{fig:vj_1_9_d1}
\end{figure}

\paragraph{Critical and weak static repulsion:} 
We consider blocking enhancement both for backtracking
and translocation with the same parameters as above but for
critical static repulsion strength and
extremely weak static repulsion, which can be realized
by clinging as mechanism that reduces backtracking
and translocation.
As expected from the general discussion above
there is no boosting. RNAP blocking enhancement
leads to jamming for all densities (Figs. \ref{fig:v_2_9_d1}
and \ref{fig:j_2_9_d1}).
This jamming is stronger
as the repulsion gets weaker, as demonstrated by the plots in 
Figs. \ref{fig:v_1_9_d1}
and \ref{fig:j_1_9_d1}.

\section{Summary and conclusions}

In this work we have studied the effect of backtracking of RNAP on the average
flux and velocity of RNAP along the DNA template during transscription elongation 
by considering the role of reverse reactions in the mechanochemical cycle that drives
translocation. As starting point we have used the mathematically tractable model
of \cite{Belitsky2019-1} which has proven to be successful in understanding the
role of interactions between RNAP in the emergence of cooperative pushing 
\cite{Galb11}, called boosting in the present work. Boosting is a macroscopic 
phenomenon that is observed in biochemical experiments
\cite{Epshtein2003-1,Epshtein2003-2,Jin10,Saek09} and signifies an
enhancement of the overall rate of transcription elongation through an increase 
of the average RNAP velocity that has its origin in pushes of stalled 
RNAP by trailing RNAP. It thus overcompensates jamming which arises from
blocking the translocation of active RNAP by stalled RNAP and thus leads
to a ``traffic jam'' \cite{Scha10} that reduces the RNAP flux and thus the
average RNAP velocity.

Significantly, as already noticed in \cite{Belitsky2019-1}, while simple 
steric excluded volume interaction between RNAP is enough to explain the 
emergence of jamming, the mere of existence of individual RNAP pushing, 
is {\it not} sufficient to explain boosting.  Likewise, it was found in the 
present work that the presence of backtracking alone (which enhances
the role of blocking) does not predict whether jamming takes over or 
whether boosting persists. Several key concepts are found to be crucial 
to understand how the interplay of microscopic forces that arise from interactions 
between individual RNAP moving along on the DNA template leads to the
macroscopic collective phenomena of jamming and boosting.

The most important (and perhaps obvious) concept is the distinction
between microscopic interactions between individual RNAP and the
emergent collective phenomena that appear on experimental macroscopic
scale. This is reflected in terminology adopted in the present paper:
On the microscopic level we speak of blocking and pushing of RNAP,
while the collective macroscopic counter parts are called jamming and
boosting. With the latter term we deviate from the more standard notion
of cooperative pushing which is what we mean by boosting, but which
somewhat obscures the fundamental distinction between microscopic
interaction between individual objects and collective outcome of this
interaction.

The second most important (and perhaps less obvious) notion is the
distinction on microscopic level between two kinds of interactions 
between individual RNAP, viz., (i) {\it static interactions} that determine 
the stationary distribution of the microscopically stochastic dynamics 
of translocation during transscription elongation, and (ii) {\it kinetic 
interactions} that determine the rates with which the various microscopic
processes in the mechanochemical cycle of RNAP translocation
occur. These two kinds of interactions are conceptually different, but both 
physically and mathematically linked: A probability distribution for the
microscopic dynamics determined by a given set of static interaction
energies cannot be stationary for just any set of transition rates
that encode the kinetic interactions. They have to be both
physically and mathematically {\it consistent}. For our model this
consistency is proved mathematically and allows us to make exact
predictions within the framework of this model. In particular, it shows
that (and how precisely) static repulsion or attraction correlate with 
kinetic blocking and pushing.

Finally, the third fundamental concept that (perhaps not surprisingly) 
plays a determining role are general characteristics of these interactions.
The point in the present context is that the process of translocation
of RNAP is permanently out of thermal equilibrium. Hence the interaction
energy appearing in the stationary distribution has to be understood
as a phenomenological {\it effective} energy. Therefore it cannot be derived 
from fundamental principles of Newtonian classical mechanics but needs to
be postulated. When developing models it makes therefore makes sense 
to be guided on the one hand by empirical data (which are usually in short
supply for the processes we have in mind) and by general theoretical notions
such as interaction range (short-ranged or long-ranged), sign (attractive or repulsive), 
and strength. As empirical data are not readily available for quantitative
predictions by the present simplified model we consider most parameters
as variables and study how the quantities that we have computed
change as these parameters are changed.

Taking these general insights as guide line, the main insight of the present
work is that  also in the presence of backtracking the strength of boosting, 
i.e., the phenomenon of cooperative pushing, is primarily determined by 
the strength of the effective static interaction between RNAP. As in the 
absence of backtracking, this static interaction needs to be repulsive and 
sufficiently strong, i.e., above a critical value that is determined by the 
interplay of the microscopic forces between two RNAP located at nearest
neighbor or next-nearest sites. If pushing is strong enough then boosting
occurs in a range of RNAP densities which is determined by the strength
of blocking due to steric excluded volume interaction and blocking
enhancement. 

This conclusion is deduced from two observations. The first point to note is 
that backtracking arising 
from the reversed mechanochemical cycle appears in the rate of 
elongation in a direct reduction of the average speed of individual RNAP 
that is determined by the rate of backward pushing and which
arises already in the absence of interaction. It is a straightforward 
consequence of the fact that occasional backtracking
reduces the average speed of a single RNAP that mostly moves by forward
translocation. Also in the interacting case this direct reduction of the 
average of an RNAP remains very small as it is shown to be proportional to 
the small bare rate of backtracking. Hence this effect has no significant
bearing on whether or not boosting occurs.

The second and more subtle manifestation of backtracking in the rate of 
elongation is in the static interaction strength itself. The consistency 
relations show that this effect is linked to the rate of pushing and blocking 
enhancement in forward translocation and hence indepedent of the overall 
bare rate of backtracking. When pushing and blocking enhancement in 
forward translocation are sufficiently strong to cause boosting, then by 
consistency also blocking enhancement of backtracking RNAP is necessarily 
large and even consistent with a very short-ranged attraction that may 
cause clinging and pulling. Thus the strength of boosting is indepedent of 
whether backtracking takes place at all.

The present model is highly stylized and in order to examine which
microscopic mechanims of backtracking affect boosting we have focussed
on a specific one. An open question to be addressed in future theoretical
work is the
role of other modes of backtracking, in particularm backtracking directly
from state 1 without PP$_i$ bound which was considered in \cite{Tripathi2008}
and  \cite{Tripathi2008} but without taking into account pushing. 
A second open problem is the range of interaction which in the present
work was taken to be at most next-nearest neighbor to allow for exact
computations in closed form. Current work on exclusion processes without
internal degree of freedom shows that the exact stationary distribution
can be constructed also for interactions with longer range \cite{Beli24}.
It is interesting to extend this approach to allow for mechanochemical
cycles with 2 or more states. This will open up the possibility to adjust
interaction parameters to less stylized interaction forces and to experimental
data that may be expected from technological advances in the observation
of motion of single RNAP.

\section*{Acknowledgements}
This work is financially supported by CAPES (Brazil)
%by Coordena\c{c}\~ao de Aperfei\c{c}oamento de Pessoal de N\'ivel Superior (CAPES) of Brazil, 
Finance Code 001, by CNPq (Brazil), grant number 140797/2018-1,
by FAPESP (Brazil), grant 2017/10555-0, and
by FCT (Portugal) through CAMGSD, IST-ID, Project UIDB/04459/2020
and by the FCT Grants 2020.03953.CEECIND and 2022.09232.PTDC.  
N. Ngoc gratefully acknowledges the financial support of CAPES and CNPq during his 
studies at the Doctorate Program in Statistics at the University of S\~ao Paulo.
%\textcolor{red}{Please check acknowledgments}

\section*{Data availability statement}

No new data were created or analysed in this study.
%\section*{Appendices}
\appendix

\section{Consistency relations and proof of the Theorem}\label{condtions_existence}

We provide a way how to find conditions on parameters appearing in the 
rates \eqref{rate1_d1}--\eqref{rate4_d1} such that the invariant measure of the 
process is of the form (\ref{invarmeas_d1}). In order to do that we employ 
the method used in \cite{Belitsky2019-1}. Namely, at equilibrium one can 
rewrite the master equation of the process in a local divergence by using a 
specific discrete form of Noether's theorem \eqref{eq21_d1}.

\subsection{Stationary condition}\label{maseq_d1}

Dividing (\ref{mastereq_d1}) by the stationary distribution 
(\ref{invarmeas_d1}), the stationary condition becomes
\begin{align}\label{stationcond_d1}
	\sum_{i=1}^{N}& \bigg[r_i(\boldsymbol{\eta}_{tlf}^i)\dfrac{\pi(\boldsymbol{\eta}_{tlf}^i)}{\pi(\boldsymbol{\eta})} + \ell_i(\boldsymbol{\eta}_{tlb}^i)\dfrac{\pi(\boldsymbol{\eta}_{tlb}^i)}{\pi(\boldsymbol{\eta})} +  a_i(\boldsymbol{\eta}_{rel}^i)\dfrac{\pi(\boldsymbol{\eta}_{rel}^i)}{\pi(\boldsymbol{\eta})} + d_i(\boldsymbol{\eta}_{bin}^i)\dfrac{\pi(\boldsymbol{\eta}_{bin}^i)}{\pi(\boldsymbol{\eta})} \nonumber \\ & -(r_i(\boldsymbol{\eta}) + \ell_i(\boldsymbol{\eta}) +  a_i(\boldsymbol{\eta}) + d_i(\boldsymbol{\eta}))\bigg] = 0.
\end{align}
Now we introduce the quantities
\begin{align}
	R_i(\boldsymbol{\eta}) & = r_i(\boldsymbol{\eta}_{tlf}^i)\dfrac{\pi(\boldsymbol{\eta}_{tlf}^i)}{\pi(\boldsymbol{\eta})} - r_i(\boldsymbol{\eta}) ,	\\
	L_i(\boldsymbol{\eta}) & = \ell_i(\boldsymbol{\eta}_{tlb}^i)\dfrac{\pi(\boldsymbol{\eta}_{tlb}^i)}{\pi(\boldsymbol{\eta})} -\ell_i(\boldsymbol{\eta}), \\
	A_i(\boldsymbol{\eta}) & = a_i(\boldsymbol{\eta}_{rel}^i)\dfrac{\pi(\boldsymbol{\eta}_{rel}^i)}{\pi(\boldsymbol{\eta})} - a_i(\boldsymbol{\eta}),\\
	D_i(\boldsymbol{\eta}) & = d_i(\boldsymbol{\eta}_{bin}^i)\dfrac{\pi(\boldsymbol{\eta}_{bin}^i)}{\pi(\boldsymbol{\eta})} - d_i(\boldsymbol{\eta}).
\end{align}

Taking into account periodicity, the stationarity condition (\ref{stationcond_d1}) is satisfied if the lattice divergence condition
\begin{equation}\label{eq21_d1}
	R_i(\boldsymbol{\eta}) + L_i(\boldsymbol{\eta}) + A_i(\boldsymbol{\eta}) + D_i(\boldsymbol{\eta}) = \Phi_{i}(\boldsymbol{\eta}) - \Phi_{i+1}(\boldsymbol{\eta})
\end{equation}
holds for all allowed configurations with a family of functions $\Phi_{i}(\boldsymbol{\eta})$ satisfying $\Phi_{N+1}(\boldsymbol{\eta}) = \Phi_{1}(\boldsymbol{\eta})$. The lattice divergence condition can be understood as a specific discrete form of Noether's theorem.

\subsection{Mapping to the headway process}

Due to steric hard core repulsion, a translocation of the $i^{th}$ rod 
from $k_{i}$ to $k_{i}+1$, corresponding to the transition $(m_{i-1}, m_i) \to 
(m_{i-1} +1, m_i - 1)$, takes place if $m_i >0$. Similarly, only if $m_{i-1}>0$ 
the 
backtracking corresponding to the transition $(m_{i-1}, m_i) \to 
(m_{i-1} -1, m_i +1)$ can occur.

In terms of the new stochastic variables $\boldsymbol{\zeta} = (\textbf{m}, 
\boldsymbol{\alpha})$ given by the distance vector $\textbf{m}$ and the state 
vector $\boldsymbol{\alpha}$ the transition rates 
(\ref{rate1_d1}--\ref{rate4_d1}) become
\begin{align}
	\tilde{r}_i(\boldsymbol{\zeta}) &= r \delta_{\alpha_i,1} (1+r^{\yesp 
	\rodr}\theta_{i-1}^0 + r^{\rodr\nop\yesp}\theta_i^1) (1-\theta_i^0);\\
	\tilde{\ell}_i(\boldsymbol{\zeta}) &= \ell \delta_{\alpha_i,2} 
	(1+\ell^{\yesp\nop\rodl}\theta_{i-1}^1 + \ell^{\rodl\yesp}\theta_i^0) 
	(1-\theta_{i-1}^0);\\
	\tilde{a}_i(\boldsymbol{\zeta}) &= a \delta_{\alpha_i,2} 
	(1+a^{\yesp\rod}\theta_{i-1}^0 + a^{\rod\yesp}\theta_i^0 + 
	a^{\yesp\rod\yesp}\theta_{i-1}^0\theta_i^0 + 
	a^{\yesp\nop\rod}\theta_{i-1}^1 + a^{\rod\nop\yesp}\theta_i^1);\\
	\tilde{d}_i(\boldsymbol{\zeta}) &=d \delta_{\alpha_i,1} 
	(1+d^{\yesp\rod}\theta_{i-1}^0 + d^{\rod\yesp}\theta_i^0 + 
	d^{\yesp\rod\yesp}\theta_{i-1}^0\theta_i^0 + 
	d^{\yesp\nop\rod}\theta_{i-1}^1 + d^{\rod\nop\yesp}\theta_i^1).
\end{align}
%In the mapping to the headway process the stationary average velocity of an RNAP is given by the stationary expectation of the function $\tilde{r}_i(\boldsymbol{\zeta}) - \tilde{\ell}_i(\boldsymbol{\zeta})$.\\

Before writing the master equation for the headway process, we introduce 
notation for the configuration that leads to a given configuration 
$\boldsymbol{\zeta}$. Namely, $\boldsymbol{\zeta}^{i-1,i}, 
\boldsymbol{\zeta}^{i,i-1}$ correspond to translocation and
backtracking respectively, and $\boldsymbol{\zeta}^{i,rel}, 
\boldsymbol{\zeta}^{i,bin} $ correspond to PP$_i$ release and binding 
respectively. Before introducing these configurations, we denote by $(k,l)$
the pair $(i-1,i)$ or $(i,i-1)$ and by $\sharp$ the superscript \textit{rel} or 
\textit{bin}. Thus, the configurations $\boldsymbol{\zeta}^{i-1,i}, 
\boldsymbol{\zeta}^{i,i-1}, \boldsymbol{\zeta}^{i,rel}$, and $ 
\boldsymbol{\zeta}^{i,bin}$
are defined by
\begin{align}
	m^{k,l}_j := m_j + \delta_{j,l} - \delta_{j,k}\ &\text{and }\  s^{k,l}_j := \alpha_j 
	+ (3-2\alpha_j) \delta_{j,i},\\
	m_j^{i,\sharp} := m_j\ &\text{and }\  \alpha_j^{i,\sharp}  := \alpha_j + 
	(3-2\alpha_j)\delta_{j,i}.
\end{align}

This yields the master equation
\begin{equation}\label{stationaryequation_d1}
	\dfrac{d\mathbb{P}(\boldsymbol{\zeta},t)}{dt} = \sum_{i=1}^{N}Q_i(\boldsymbol{\zeta},t)
\end{equation}
with
\begin{align}
	\begin{split}
		Q_i(\boldsymbol{\zeta},t) = & \ \  \tilde{r}_i(\boldsymbol{\zeta}^{i-1,i})\mathbb{P}(\boldsymbol{\zeta}^{i-1,i},t) - \tilde{r}_i(\boldsymbol{\zeta})\mathbb{P}(\boldsymbol{\zeta},t) 
		+ \tilde{\ell}_i(\boldsymbol{\zeta}^{i,i-1})\mathbb{P}(\boldsymbol{\zeta}^{i,i-1},t) - \tilde{\ell}_i(\boldsymbol{\zeta})\mathbb{P}(\boldsymbol{\zeta},t)\\  
		& + \tilde{a}_i(\boldsymbol{\zeta}^{i,rel})\mathbb{P}(\boldsymbol{\zeta}^{i,rel},t) - \tilde{a}_i(\boldsymbol{\zeta})\mathbb{P}(\boldsymbol{\zeta},t) +
		\tilde{d}_i(\boldsymbol{\zeta}^{i,bin})\mathbb{P}(\boldsymbol{\zeta}^{i,bin},t) - \tilde{d}_i(\boldsymbol{\zeta})\mathbb{P}(\boldsymbol{\zeta},t)
	\end{split}
\end{align}
where
\begin{align}
	\tilde{r}_i(\boldsymbol{\zeta}^{i-1,i}) &= r \delta_{\alpha_i,2} (1+r^{\yesp 
	\rodr}\theta_{i-1}^1 + r^{\rodr\nop\yesp}\theta_i^0) (1-\theta_{i-1}^0),\\
	\tilde{\ell}_i(\boldsymbol{\zeta}^{i,i-1}) &= \ell \delta_{\alpha_i,1} 
	(1+\ell^{\yesp\nop\rodl}\theta_{i-1}^0 + \ell^{\rodl\yesp}\theta_i^1) 
	(1-\theta_{i}^0),\\
	\tilde{a}_i(\boldsymbol{\zeta}^{i,rel})&= a \delta_{\alpha_i,1} 
	(1+a^{\yesp\rod}\theta_{i-1}^0 + a^{\rod\yesp}\theta_i^0 + 
	a^{\yesp\rod\yesp}\theta_{i-1}^0\theta_i^0 + 
	a^{\yesp\nop\rod}\theta_{i-1}^1 + a^{\rod\nop\yesp}\theta_i^1),\\
	\tilde{d}_i(\boldsymbol{\zeta}^{i,bin}) &=d \delta_{\alpha_i,2} 
	(1+d^{\yesp\rod}\theta_{i-1}^0 + d^{\rod\yesp}\theta_i^0 + 
	d^{\yesp\rod\yesp}\theta_{i-1}^0\theta_i^0 + 
	d^{\yesp\nop\rod}\theta_{i-1}^1 + d^{\rod\nop\yesp}\theta_i^1).
\end{align}

Thanks to the discrete version of Noether theorem, one can rephrase the 
stationarity condition for the headway process in a local divergence form 
which is equivalent to \eqref{eq21_d1}. Let us first introduce the following 
notations
\begin{align}
	\tilde{R}_i(\boldsymbol{\zeta}) &= \tilde{r}_i(\boldsymbol{\zeta}^{i-1,i})\dfrac{\tilde{\pi}(\boldsymbol{\zeta}^{i-1,i})}{\tilde{\pi}(\boldsymbol{\zeta})} -\tilde{r}_i(\boldsymbol{\zeta}),\\
	\tilde{L}_i(\boldsymbol{\zeta}) &= \tilde{\ell}_i(\boldsymbol{\zeta}^{i,i-1})\dfrac{\tilde{\pi}(\boldsymbol{\zeta}^{i,i-1})}{\tilde{\pi}(\boldsymbol{\zeta})} -\tilde{\ell}_i(\boldsymbol{\zeta}),\\
	\tilde{A}_i(\boldsymbol{\zeta}) &= \tilde{a}_i(\boldsymbol{\zeta}^{i,rel})\dfrac{\tilde{\pi}(\boldsymbol{\zeta}^{i,rel})}{\tilde{\pi}(\boldsymbol{\zeta})} -\tilde{a}_i(\boldsymbol{\zeta}),\\
	\tilde{D}_i(\boldsymbol{\zeta}) &= \tilde{d}_i(\boldsymbol{\zeta}^{i,bin})\dfrac{\tilde{\pi}(\boldsymbol{\zeta}^{i,bin})}{\tilde{\pi}(\boldsymbol{\zeta})} -\tilde{d}_i(\boldsymbol{\zeta}).
\end{align}
Again, we make use of $(k,l)$ for $(i-1,i)$ or $(i,i-1)$ and $\sharp$ for the superscript $rel$ or $bin$. Notice that
\begin{align}
	&\theta_j^p(\boldsymbol{\zeta}^{k,l}) = 
	\delta_{m_j+\delta_{j,l}-\delta_{j,k},p} = 
	\theta_j^{p-\delta_{j,l}+\delta_{j,k}}(\boldsymbol{\zeta}) \text{ and } 
	\delta_{\alpha_i^{k,l},\alpha} =\delta_{\alpha_i,3-\alpha};\\
	&\theta_j^p(\boldsymbol{\zeta}^{i,\sharp}) = 
	\theta_j^{p}(\boldsymbol{\zeta}) \text{ and } 
	\delta_{\alpha_i^{i,\sharp},\alpha} =\delta_{\alpha_i,3-\alpha},
\end{align}
%\begin{align}
%	&\theta_j^p(\boldsymbol{\zeta}^{i-1,i}) = 
%\delta_{m_j+\delta_{j,i}-\delta_{j,i-1},p} = 
%\theta_j^{p-\delta_{j,i}+\delta_{j,i-1}}(\boldsymbol{\zeta}) \text{ and } 
%\delta_{\alpha_i^{i-1,i},\alpha} =\delta_{\alpha_i,3-\alpha};\\
%	&\theta_j^p(\boldsymbol{\zeta}^{i,i-1}) = 
%\delta_{m_j-\delta_{j,i}+\delta_{j,i-1},p} = 
%\theta_j^{p+\delta_{j,i}-\delta_{j,i-1}}(\boldsymbol{\zeta}) \text{ and } 
%\delta_{\alpha_i^{i,i-1},\alpha} =\delta_{\alpha_i,3-\alpha};\\
%	&\theta_j^p(\boldsymbol{\zeta}^{i,rel}) = \theta_j^{p}(\boldsymbol{\zeta}) 
%\text{ and } \delta_{\alpha_i^{i,rel},\alpha} =\delta_{\alpha_i,3-\alpha};\\
%	&\theta_j^p(\boldsymbol{\zeta}^{i,bin}) = \theta_j^{p}(\boldsymbol{\zeta}) 
%\text{ and } \delta_{\alpha_i^{i,bin},\alpha} =\delta_{\alpha_i,3-\alpha}.
%\end{align}
so that one gets
\begin{align}
	\dfrac{\tilde{\pi}(\boldsymbol{\zeta}^{i,\sharp})}{\tilde{\pi}(\boldsymbol{\zeta})}
	 &=x^{3-2\alpha_i},\\
	\dfrac{\tilde{\pi}(\boldsymbol{\zeta}^{k,l})}{\tilde{\pi}(\boldsymbol{\zeta})} & 
	= x^{-3+2\alpha_i}y^{\theta_{k}^0 -\theta_{k}^1+\theta_{l}^0}.
\end{align}
Hence,
%\begin{equation}
%	\dfrac{\tilde{\pi}(\boldsymbol{\zeta}^{i,rel})}{\tilde{\pi}(\boldsymbol{\zeta})} 
%=x^{3-2\alpha_i},
%\end{equation}
%\begin{equation}
%	\dfrac{\tilde{\pi}(\boldsymbol{\zeta}^{i,bin})}{\tilde{\pi}(\boldsymbol{\zeta})} 
%=x^{3-2\alpha_i},
%\end{equation}
%\begin{align}
%	\dfrac{\tilde{\pi}(\boldsymbol{\zeta}^{i-1,i})}{\tilde{\pi}(\boldsymbol{\zeta})} 
%& =\dfrac{y^{-\theta_{i-1}^0(\boldsymbol{\zeta}^{i-1,i})} 
%x^{-3/2+\alpha_{i-1}(\boldsymbol{\zeta}^{i-1,i})}}{y^{-\theta_{i-1}^0} 
%x^{-3/2+\alpha_{i-1}}}\cdot \dfrac{y^{-\theta_i^0(\boldsymbol{\zeta}^{i-1,i})} 
%x^{-3/2+\alpha_i(\boldsymbol{\zeta}^{i-1,i})}}{y^{-\theta_{i}^0} 
%x^{-3/2+\alpha_i}} \nonumber\\
%	& = x^{-1}y^{\theta_{i-1}^0 +\theta_{i}^0-\theta_{i-1}^1},
%\end{align}
%%and
%\begin{align}
%	\dfrac{\tilde{\pi}(\boldsymbol{\zeta}^{i,i-1})}{\tilde{\pi}(\boldsymbol{\zeta})} 
%& =\dfrac{y^{-\theta_{i-1}^0(\boldsymbol{\zeta}^{i,i-1})} 
%x^{-3/2+\alpha_{i-1}(\boldsymbol{\zeta}^{i,i-1})}}{y^{-\theta_{i-1}^0} 
%x^{-3/2+\alpha_{i-1}}}\cdot \dfrac{y^{-\theta_i^0(\boldsymbol{\zeta}^{i,i-1})} 
%x^{-3/2+\alpha_i(\boldsymbol{\zeta}^{i,i-1})}}{y^{-\theta_{i}^0} 
%x^{-3/2+\alpha_i}} \nonumber \\
%	& = xy^{\theta_{i}^0 -\theta_{i}^1+\theta_{i-1}^0  }.
%\end{align}
\begin{align}
	\tilde{R}_i(\boldsymbol{\zeta}) =&\  x^{-1}y^{\theta_{i-1}^0 
	+\theta_{i}^0-\theta_{i-1}^1}  r \delta_{\alpha_i,2} (1+r^{\yesp 
	\rodr}\theta_{i-1}^1 + r^{\rodr\nop\yesp}\theta_i^0) (1-\theta_{i-1}^0) 
	\nonumber\\
	& - r \delta_{\alpha_i,1} (1+r^{\yesp\rodr}\theta_{i-1}^0 + 
	r^{\rodr\nop\yesp}\theta_i^1) (1-\theta_i^0),\\
	\tilde{L}_i(\boldsymbol{\zeta}) =&\ xy^{\theta_{i}^0 
	-\theta_{i}^1+\theta_{i-1}^0  }  \ell \delta_{\alpha_i,1} 
	(1+\ell^{\yesp\nop\rodl}\theta_{i-1}^0 + \ell^{\rodl\yesp}\theta_i^1) 
	(1-\theta_{i}^0) \nonumber\\
	&- \ell \delta_{\alpha_i,2} (1+\ell^{\yesp\nop\rodl}\theta_{i-1}^1 + 
	\ell^{\rodl\yesp}\theta_i^0) (1-\theta_{i-1}^0),\\
	\tilde{A}_i(\boldsymbol{\zeta}) =&\  (x\delta_{\alpha_i,1} 
	-\delta_{\alpha_i,2})\kappa^{\rod}(1+a^{\yesp\rod}\theta_{i-1}^0 + 
	a^{\rod\yesp}\theta_i^0 + a^{\yesp\rod\yesp}\theta_{i-1}^0\theta_i^0 + 
	a^{\yesp\nop\rod}\theta_{i-1}^1 + a^{\rod\nop\yesp}\theta_i^1),\\
	\tilde{D}_i(\boldsymbol{\zeta}) = &\  (x^{-1}\delta_{\alpha_i,2} 
	-\delta_{\alpha_i,1})\tau^{\rod} (1+d^{\yesp\rod}\theta_{i-1}^0 + 
	d^{\rod\yesp}\theta_i^0 + d^{\yesp\rod\yesp}\theta_{i-1}^0\theta_i^0 + 
	d^{\yesp\nop\rod}\theta_{i-1}^1 + d^{\rod\nop\yesp}\theta_i^1).
\end{align}
One requires
\begin{equation}\label{Noether_d1}
	\tilde{R}_i + \tilde{L}_i + \tilde{A}_i + \tilde{D}_i = \tilde{\Phi}_{i-1} - \tilde{\Phi}_{i},
\end{equation}
where $\tilde{\Phi}_i$ is of the form $\tilde{\Phi}_i = (e+ f\theta_{i}^0 + 
h\theta_{i}^1)(\delta_{\alpha_i,1}+\delta_{\alpha_i,2}) = e+ f\theta_{i}^0 + 
h\theta_{i}^1$. Notice that $\tilde{\Phi}_i$ must be of that form since 
$\tilde{R}_i, \tilde{L}_i, \tilde{A}_i, \tilde{D}_i$ depend on the state of rod $i$ 
and variables $\theta_{i-1}^0, \theta_{i-1}^1, \theta_{i}^0, \theta_{i}^1$ 
belonging to $\{0,1\}$.

By considering all possible cases of \eqref{Noether_d1}, one first gets
\begin{align}
	f & =\dfrac{-r (1+ r^{\yesp\rodr})+x\ell(1+\ell^{\rodl\yesp})}{1+x} 
	\label{value_b}\\
	h & = \dfrac{r r^{\rodr\nop\yesp} - x\ell \ell^{\yesp\nop\rodl} }{1+x} 
	\label{value_h}
\end{align}
 and then one gets the stationary conditions \eqref{stationarycondition1_d1_1}--\eqref{end_d1_1}. However, we shall give a short proof of the above claim in the next subsection after knowing the results.

%\section{Proofs }
\subsection{Proof of Theorem \ref{main_theorem}}
As previously noted, it is necessary to account for all instances of \eqref{Noether_d1} in order to identify the constraints. Yet, this task may seem tedious in ensuring the accuracy of the results. In this context, we aim to present a concise proof of Theorem \ref{main_theorem}. Consequently, we will demonstrate that given conditions (\ref{stationarycondition1_d1_1})--(\ref{end_d1_1}), the process's invariant measure adheres to the structure described in \eqref{main_theorem_d1}. Our objective is achieved by confirming that \eqref{Noether_d1} holds for all configurations. 

\begin{itemize}
	\item If $m_{i-1}, m_{i} > 1$, one has $	\tilde{R}_i  = 
	x^{-1}r\delta_{\alpha_i,2} -r\delta_{\alpha_i,1}, \tilde{L}_i = 
	x\ell\delta_{\alpha_i,1} -\ell\delta_{\alpha_i,2},
	\tilde{A}_i  = (x\delta_{\alpha_i,1} -\delta_{\alpha_i,2})a, \tilde{D}_i  = 
	(x^{-1}\delta_{\alpha_i,2} -\delta_{\alpha_i,1})d$. Notice that in this case, the 
	left-hand side of \eqref{Noether_d1} is 0. If the state of rod is 1 meaning that 
	$\alpha_i =1$, one has $\tilde{R}_i + \tilde{L}_i + \tilde{A}_i + \tilde{D}_i = 
	-r + x\ell + xa -d$ which is 0 due to \eqref{stationarycondition1_d1_1}. 
	Similarly, for the case $\alpha_i =2$, \eqref{Noether_d1} holds.
	\item If $m_{i-1}>1, m_{i} = 1$, one has $	\tilde{R}_i  = 
	x^{-1}r\delta_{\alpha_i,2} -r\delta_{\alpha_i,1}(1+r^{\rodr\nop\yesp}), 
	\tilde{L}_i = xy^{-1}\ell\delta_{\alpha_i,1}(1+\ell^{\rodl\yesp}) 
	-\ell\delta_{\alpha_i,2},
	\tilde{A}_i  = (x\delta_{\alpha_i,1} 
	-\delta_{\alpha_i,2})a(1+a^{\rod\nop\yesp}), \tilde{D}_i  = 
	(x^{-1}\delta_{\alpha_i,2} -\delta_{\alpha_i,1})d(1+d^{\rod\nop\yesp})$. 
	Notice that in this case the left-hand side of \eqref{Noether_d1} is $-h$. If 
	$\alpha_i =1$, one has $\tilde{R}_i + \tilde{L}_i + \tilde{A}_i + \tilde{D}_i = 
	-r(1+r^{\rodr\nop\yesp}) + xy^{-1}\ell(1+\ell^{\rodl\yesp}) + 
	xa(1+a^{\rod\nop\yesp}) -d(1+d^{\rod\nop\yesp})$. It is easy to check from 
	\eqref{stationarycondition1_d1_1}, \eqref{stationarycondition1_d1_2}, and 
	\eqref{end_d1_1} that $\tilde{R}_i + \tilde{L}_i + \tilde{A}_i + \tilde{D}_i  = -h$ 
	where $h$ is defined in \eqref{value_h}. Thus, \eqref{Noether_d1} holds for 
	this case. Similarly, for the case $\alpha_i =2$, \eqref{Noether_d1} holds as 
	well.
	\item For the rest of the cases: $m_{i-1}>1, m_{i} = 0$; $m_{i-1}=1, m_{i} > 1$; $m_{i-1}=0, m_{i} > 1$; $m_{i-1}=1, m_{i} = 1$; $m_{i-1}=1, m_{i} = 0$; $m_{i-1}=0, m_{i} = 1$; $m_{i-1}=0, m_{i} = 0$, one considers similarly to show that \eqref{Noether_d1} holds.
\end{itemize}
The proof is complete.

\end{document}